%% file: spgemm.tex
\documentclass[conference]{IEEEtran}
\usepackage{tikz}
\usepackage{url}
\usepackage{pdfpages}
\usepackage{afterpage}
\usepackage{color, colortbl}
\usepackage{graphicx}
\usepackage{amsmath}
\usepackage{amsfonts}
\usepackage{import}
\usepackage{pifont}
\usepackage{multirow}
\usepackage{subfigure}
\usepackage{footmisc}
\usepackage{algorithm}
\usepackage[noend]{algpseudocode}
\usepackage{amssymb}
\usepackage[
separate-uncertainty = true,
multi-part-units = repeat
]{siunitx}

\def\BState{\State\hskip-\ALG@thistlm}

\definecolor{Gray}{gray}{0.9}
\definecolor{LightCyan}{rgb}{0.88,1,1}
\pdfoutput=1
\graphicspath {{figures/}}

\makeatletter
\DeclareRobustCommand*\textsubscript[1]{%
	\@textsubscript{\selectfont#1}}
\def\@textsubscript#1{%
	{\m@th\ensuremath{_{\mbox{\fontsize\sf@size\z@#1}}}}}
\makeatother

\newcommand\ma[1]{\textbf{\textit{#1}}}

\usepackage[flushleft]{threeparttable}

\begin{document}
	
	\title{Efficient Sparse-Dense Matrix-Matrix Multiplication on GPUs Using the Customized Sparse Storage Format}

	\author{\IEEEauthorblockN{Shaohuai Shi, Qiang Wang, Xiaowen Chu}
		\IEEEauthorblockA{Department of Computer Science, Hong Kong Baptist University
			\\\{csshshi, qiangwang, chxw\}@comp.hkbu.edu.hk}
	}
		\maketitle
\begin{abstract}
Multiplication of a sparse matrix to a dense matrix (SpDM) is widely used in many areas like scientific computing and machine learning. However, existing works under-look the performance optimization of SpDM on modern many-core architectures like GPUs. The storage data structures help sparse matrices store in a memory-saving format, but they bring difficulties in optimizing the performance of SpDM on modern GPUs due to irregular data access of the sparse structure, which results in lower resource utilization and poorer performance. In this paper, we refer to the roofline performance model of GPUs to design an efficient SpDM algorithm called GCOOSpDM, in which we exploit coalescent global memory access, fast shared memory reuse and more operations per byte of global memory traffic. Experiments are evaluated on three Nvidia GPUs (i.e., GTX 980, GTX Titan X Pascal and Tesla P100) with CUDA-8.0 using a large number of matrices including a public dataset and randomly generated matrices. Experimental results show that GCOOSpDM achieves 1.5-8$\times$ speedup over  Nvidia's library cuSPARSE in many matrices. We also analyze instruction-level operations on a particular GPU to understand the performance gap between GCOOSpDM and cuSPARSE. The profiled instructions confirm that cuSPARSE spends a lot of time on slow memory access (including DRAM access and L2 cache access), while GCOOSpDM transfers such slow memory access to faster shared memory, which mainly contributes to the performance gain. Results also show that GCOOSpDM would outperform the dense algorithm (cuBLAS) with lower sparsity than cuSPARSE on GPUs.
\end{abstract}


\section{Introduction}
Sparse-dense matrix-matrix multiplication (SpDM) has many application areas. It is not only exploited in traditional research fields (e.g., graph analytics \cite{tiskin2001all}, biology \cite{vazquez2010matrix}), but becoming a potential faster implementation for sparse deep learning \cite{liu2015sparse}\cite{shi2017speeding}\cite{wen2017learning}\cite{sun2017meprop}\cite{shi2019distributed}. However, it requires very high sparsity of the model to achieve accelerated speed compared to the original dense implementations \cite{narang2017exploring}.

Dense matrix multiplication, i.e., $C=A\times B$ or general purpose matrix multiplication (GEMM) has been well studied on GPUs to achieve high efficiency \cite{volkov2008benchmarking}\cite{chu2009practical}\cite{nath2010improved}\cite{matsumoto2011multi}\cite{kurzak2012autotuning}\cite{lai2013performance}\cite{abdelfattah2016performance}\cite{zhang2017understanding}\cite{yan2020demystifying}\cite{liu2018g}. However, multiplication of a sparse matrix to a dense matrix (SpDM), in which the sparse matrix is stored with memory-saving formats like compressed row storage (CRS) \cite{dongarra2000compressed}, is understudied, and it easily loses efficiency on modern GPUs. For example, the time cost of calculating the multiplication of a $8000\times 8000$ sparse matrix with sparsity of $0.9$ (i.e., $90\%$ of elements are zeros) to a dense matrix with single precision requires $780ms$ by using cuSPARSE on an Nvidia Tesla P100 GPU, while the corresponding dense algorithm by cuBLAS only requires $121ms$.\footnote{Both cuSPARSE and cuBLAS are from the library of CUDA-8.0.} In other words, though the sparse matrix can reduce the number of multiplication and accumulation operations (MACs) by $90\%$ (since a zero element times any numbers produces zeros that has no contribution to the final results, so such operations can be avoided.), the highly optimized cuBLAS is about $7\times$ faster than cuSPARSE in the above example. For a much higher sparsity of $0.995$, cuSPARSE can be about $50\%$ faster than cuBLAS at the dimension of $8000\times 8000$ matrices on the P100 GPU. High sparsity requirement on SpDM makes it difficult to be deployed as the efficient implementation of matrix multiplication because of the inefficient algorithm design of the SpDM algorithm in cuSPARSE. In practical problems, on one hand, if the sparsity is not high enough, doing SpDM could result in very low efficiency, while using the dense form could get results faster if there is enough memory; on the other hand, if the sparsity is very high, using the dense form not only leads to low efficiency, but it also wastes memory. From our empirical studies of cuSPARSE and cuBLAS, the sparse algorithm of cuSPARSE requires the matrix sparsity to be larger than $0.99$ to outperform the dense counterpart of cuBLAS. One of our key observations of cuSPARSE is that it has many slow memory access that easily leaves the computational resources (i.e., cores) stale in its SpDM APIs. To this end, we would like to design an efficient SpDM algorithm to better utilize the GPU computational resources.

Only a small number of research works focus on high-performance SpDM algorithms for modern GPUs. The most relevant work is \cite{ortega2013fastspmm}, \cite{yang2018design} and \cite{parger2020speck}\cite{jiang2020novel}. On one hand, Ortega et al. \cite{ortega2013fastspmm} try to better optimize the GPU memory access pattern (i.e., coalesced memory access) to achieve higher efficiency. On the other hand, besides the optimization of coalesced memory access, Yang et al. \cite{yang2018design} use the principles of row split \cite{bell2009implementing} and merge path \cite{merrill2016merge} in sparse matrix-dense vector multiplication (SpMV) to design more efficient algorithms for SpDM on GPUs. Jiang et al. \cite{jiang2020novel} mainly re-order the row data and Parger et al. \cite{parger2020speck} propose the parameter tuning technique to optimize the performance of SpDM. However, in \cite{yang2018design}, the authors design their algorithms mainly for the cases that the dense matrices are tall-skinny, and it requires a heuristic to choose whether to use merge-based or row split for better performance. In this paper, we not only exploit the GPU algorithm optimization principles (e.g., coalesced memory access), but also revisit the popular roofline performance model \cite{williams2009roofline} on GPUs to analyze how to increase operational intensity, and then we propose an efficient SpDM algorithm. Our contributions are summarized as follows:

\begin{itemize}
	\item We design an efficient SpDM algorithm called GCOOSpDM on GPUs with several optimization techniques including coalescing memory access, bank conflict avoidance of the shared memory and high computation-to-memory ratios.
	\item We evaluate the proposed algorithm on a large number of sparse matrices including the public dataset and randomly generated matrices, and the experimental results show that GCOOSpDM outperforms cuSPARSE 1.5-8$\times$ faster in a large proportion of matrices on Nvidia GPUs.
	\item We conduct instruction-level analysis for the kernels of GCOOSpDM and cuSPARSE, and the profiled results confirm that our proposed algorithm uses much less slow memory access (DRAM and L2 cache) than cuSPARSE.
	\item As compared to cuSPARSE, GCOOSpDM decreases the sparsity requirement from $0.99$ to $0.98$ in order to outperform dense implementation of cuBLAS.
\end{itemize}

The rest of the paper is organized as follows. Section \ref{section:preliminaries} gives introductions to the preliminaries related to SpDM and GEMM. We present our proposed algorithm for efficient SpDM in Section \ref{section:algorithm}. The experimental evaluation and analysis are illustrated in Section \ref{section:evaluation}. Section \ref{section:relatedwork} introduces the related work, and finally we conclude this paper in Section \ref{section:conclusion}.

\section{Preliminaries}\label{section:preliminaries}
A multiplication of two matrices $\ma{A}\in \mathbb{R}^{m\times k}$ and $\ma{B}\in \mathbb{R}^{k\times n}$ produces an result matrix $\ma{C}\in \mathbb{R}^{m\times n}$, i.e.,
\begin{equation}
\ma{C}(i,j)=\sum_{l=0}^{l=k-1}\ma{A}(i,l)\times \ma{B}(l, j).
\end{equation}
To simplify the analysis of the algorithms, we assume that the dimensions of $\ma{A}$ and $\ma{B}$ are both $n\times n$. The sparsity $s$ of matrix $\ma{A}$ is defined as the ratio of the number of zero elements over the total number of elements.

\subsection{The roofline model}
\begin{figure}[!h]
		\centering
		\includegraphics[width=0.8\linewidth]{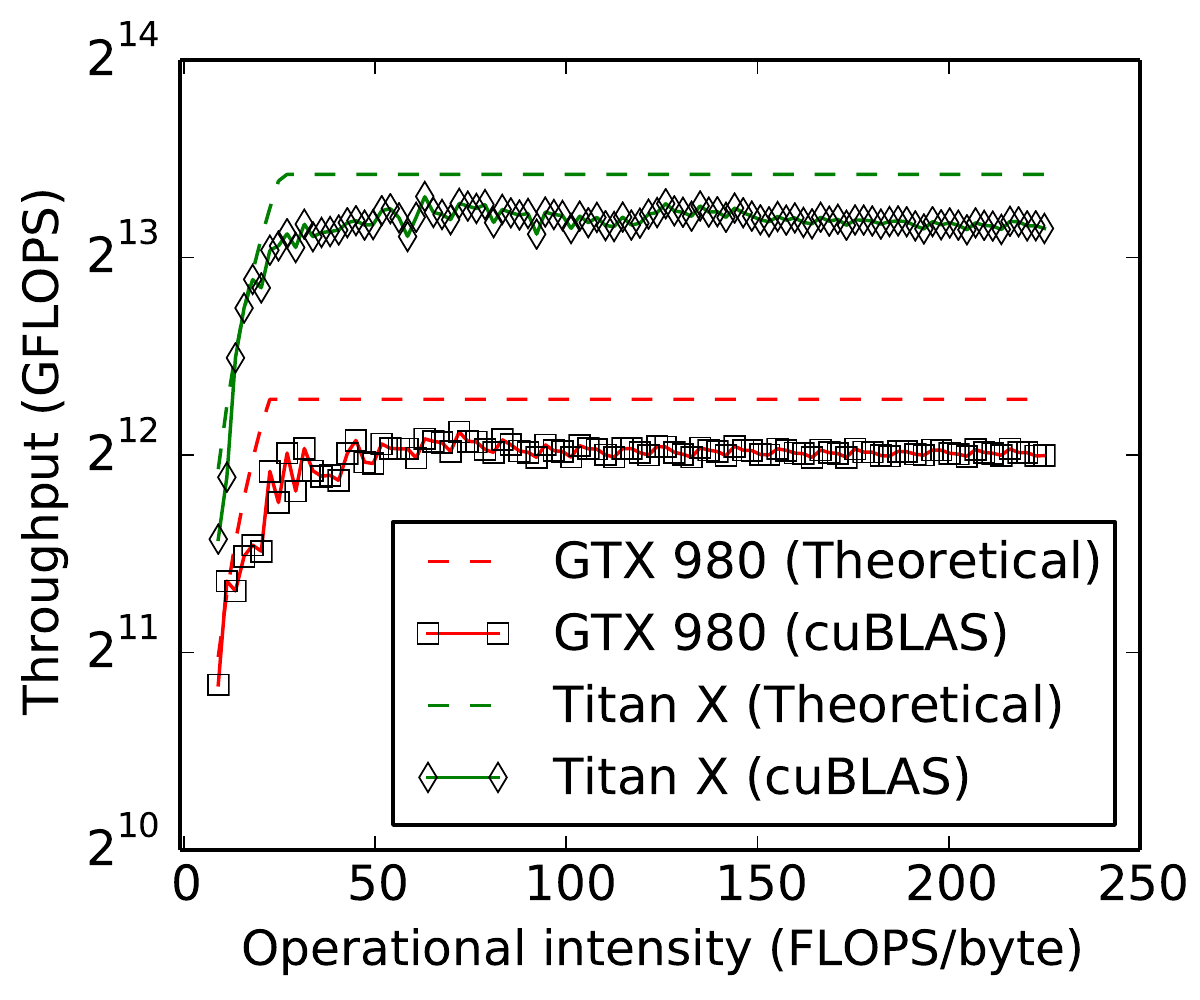}
		\caption{The roofline models for theoretical peak throughput and cuBLAS throughput with single-precision on GPUs.}
		\label{fig:gpucublas}
\end{figure}

The roofline model \cite{williams2009roofline} is commonly used in performance modeling of multi-core/many-core architectures like GPUs \cite{kim2011performance}\cite{zhang2017understanding}\cite{konstantinidis2017quantitative}. The term operational intensity $r$ (operations per byte of DRAM traffic) is defined to predict the performance of kernels. In the model, there is an upper bound of the GPU throughput when $r$ reaches some threshold, which indicates the program is computation-bound. If $r$ is smaller than the threshold, the GPU throughput is a linear function with respect to $r$, which indicates the program is memory-bound. Using cuBLAS GEMM as an example, in Fig. \ref{fig:gpucublas}, we compare the experimental throughput of dense matrix multiplication with the theoretical throughput from roofline model on two different Nvidia GPUs, GTX980 and Titan X. 

Though GEMM in cuBLAS has achieved nearly optimal throughput on matrix multiplication, directly applying GEMM for sparse matrices could result in many useless calculations due to the large amount of zeros. The irregular non-zero elements in sparse matrices make the data access from global memory to registers become the bottleneck of matrix multiplication. In other words, each time of data reading from the sparse matrix, only a limited number of computational operations. Therefore, algorithms for SpDM are generally memory-bound, and for such problems, one should design the algorithm to increase $r$ to achieve higher efficiency. 

\subsection{GPU memory hierarchy} 
From the roofline model, one should improve the memory access efficiency to fully utilize the computational power of GPUs. There are several types of memories in the GPU memory hierarchy. From fast to slow of access speed, it contains registers, the shared memory (or L1 cache), L2 cache and the global memory \cite{volkov2008benchmarking}\cite{mei2017dissecting}\cite{mei2014benchmarking}. The shared memory and  global memory are two kinds of memories that can be flexibly manipulated by programming. In general, data that is repeatedly used could be put into the shared memory or registers for better utilization of GPU cores.

\subsection{COO: The coordinate storage format}
Assume that the matrix is a row-major matrix. The coordinate storage format (COO) \cite{bell2009implementing} is a simple storage scheme for sparse matrices. COO uses an array $values$ to store the values of all non zero elements. The coordinate information of each non zero element is sequentially stored in array $rows$ and array $cols$ respectively. Take a real-valued example of a $4\times 4$ sparse matrix as follows: 
\[
\ma{A}=
\begin{bmatrix}
	7 & 0 & 0 & 8 \\
	0 & 10 &0 & 0 \\
	9 & 0 & 0 & 0 \\
	0 & 0 & 6 & 3
\end{bmatrix},
\]
the COO format of $\ma{A}$ is represented by
\begin{align*}
values&=[7,8,10,9,6,3],\\
rows &=[0,0,1,2,3,3],\\
cols &=[0,3,1,0,2,3].
\end{align*}

\section{Efficient Algorithm Design}\label{section:algorithm}
In this section, we describe the design of our proposed efficient SpDM algorithm on GPUs including the customized storage format for sparse matrices and its conversion from the dense ones. According to the above analysis in operations of SpDM on GPUs, we first design a new sparse format called grouped COO (GCOO), which is convenient for coalesced memory access and is useful to increase the operational intensity $r$. Then we propose an efficient SpDM algorithm by using GCOO. 

\subsection{GCOO: Grouped COO storage format}
A similar format of GCOO is the sliced COO (SCOO) format proposed in \cite{dang2012sliced}, with which the authors achieved higher throughput on sparse matrix-vector multiplication (SpMV) on both CPUs and GPUs. In this paper, we bring the idea of SCOO to propose GCOO for matrix multiplication. The sparse matrix is partitioned to $g$ groups according to the number of columns $n$, and each group is stored in the COO format, so we call it GCOO. For an $n\times n$ matrix stored in the GCOO format, there are $g=\lfloor \frac{n+p-1}{p} \rfloor$ groups, and each group contains $p$ columns except the last one who has $n-(g-1)\times p$ columns. If $n$ is divisible by $p$, then the last group also has $p$ columns. In GCOO, each group is stored in the COO format, and COOs from all groups are concatenated into one array. Let group $i$ be stored in the COO format with $rows_i$, $cols_i$ and $values_i$, where $i=0,1,...,g-1$. We have the stored values of GCOO with $rows$, $cols$ and $values$ that are generated from the concatenation of $rows_i$, $cols_i$ and $values_i$ respectively.

\begin{figure}[!h]
	\centering
	\includegraphics[width=0.4\linewidth]{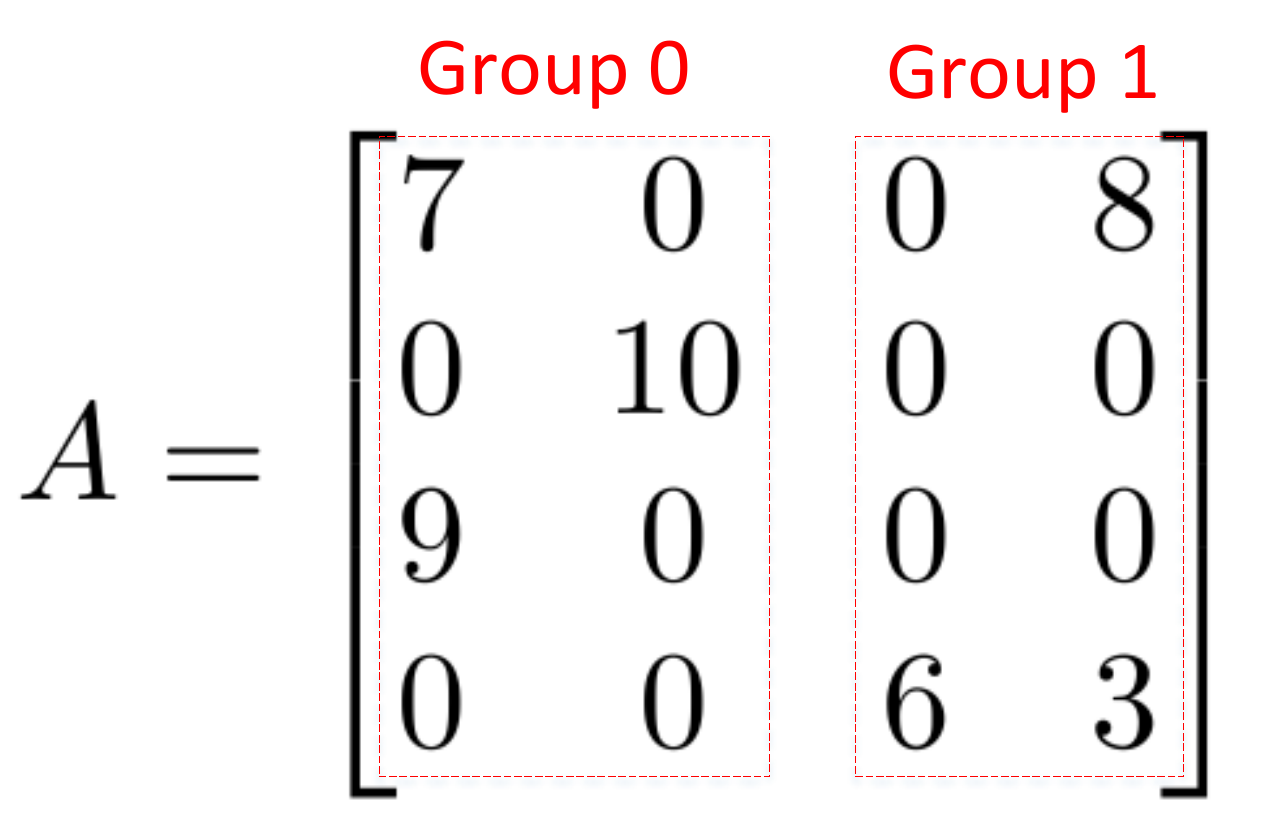}
	\caption{An example of GCOO. It has 2 groups, and each group contains 2 columns (i.e., $p=2$).}
	\label{fig:gcoo}
\end{figure} 
An example of grouping in matrix $\ma{A}$ is shown in Fig. \ref{fig:gcoo}. Matrix $\ma{A}$ is divided into to 2 groups. Group $0$ is represented by $rows_0=[0,1,2]$, $cols_0=[0,1,0]$ and $values_0=[7,10,9]$; and group $1$ is represented by $rows_1=[0,3,3]$, $cols_1=[3,2,3]$ and $values_1=[8,6,3]$. Finally, two groups are concatenated into one array with an extra index array $gIdxes$ to indicate which positions are corresponding to related groups. Therefore, the final stored format of GCOO is as follows:
\begin{align*}
values&=[7,10,9,8,6,3],\\
rows &=[0,1,2,0,3,3],\\
cols &=[0,1,0,3,2,3],\\
gIdxes &=[0,3],
\end{align*}
where $gIdxes$ is an auxiliary array to store the group indexes. It is noted that $rows$, $cols$ and $values$ in GCOO are not the same as those of COO since a single group in GCOO is in a COO format. In order to easily access each group's elements, we use an extra auxiliary array, $nnzPerGroup$, to store the number of non-zero elements in each group. In the above example, the values of $nnzPerGroup$ should be:
\begin{align*}
nnzPerGroup &=[3,3].
\end{align*}
In practice, GCOO spends slightly more memory space than COO and CSR, but it provides more convenient access of data with a higher probability. The comparison of memory consumption to store an $n\times n$ matrix with a sparsity of $s$ (note that $nnz=s\times n^2$) is shown in Table \ref{table:memcon}.
\begin{table}[!ht]
	\centering
	\caption{Memory consumption of different formats.}
	\label{table:memcon}
	\begin{tabular}{|l|l|}
		\hline
		Format& 	Memory complexity \\\hline
		\hline
		CSR & $2\times nnz+n$  \\\hline
		COO &  $3\times nnz$ \\\hline
		GCOO & $3\times nnz + 2\times \lfloor \frac{n+p-1}{p} \rfloor$\\\hline
	\end{tabular}
\end{table}

The main advantage of GCOO is to help reuse the data from slow memories (e.g., global memory and L2 cache). Specifically, if there exist two or more continuous non-zero elements in one group that are in the same row, then the fetched element from the dense matrix $\ma{B}$ can be reused in the register instead of being read from the slow memory again.

\subsection{Matrix conversion to GCOO}
For the cases that the input matrices $\ma{A}$ and $\ma{B}$ are stored in the dense form, there would be an extra overhead in the format conversion to apply the SpDM algorithm. For example, cuSPARSE provides an API ``cusparseSdense2csr'' to convert the dense matrix to the CSR format so that one can apply the SpDM APIs. For our proposed GCOO, we also need to provide an efficient conversion scheme to convert the dense matrix to GCOO. We use two steps to convert the dense matrix to the GCOO storage.

Step 1: Count the number of non-zero elements. To convert a dense form of a matrix to the sparse form, one should first count the number of non-zero elements  ($nnz$) of that matrix in order to allocate the memory according to the value of $nnz$. As for GCOO, we have pre-grouped the matrix by pre-defined $p$, so it is straightforward to calculate the non-zero elements in parallel for different groups such that the array $nnzPerGroup$ can also be calculated. Therefore, in this step, $nnz$, $gIdxes$ and $nnzPerGroup$ can be calculated by scanning the original dense matrix.

Step 2: Store the non-zero elements to $rows$, $cols$ and $values$. First, the memories of $rows$, $cols$ and $values$ are allocated according to $nnz$, and then we can read the non-zero elements with their coordinate information and write them to $rows$, $cols$, and $values$ according to the indexes by $nnzPerGroup$ in parallel. 

The pseudocode of the matrix conversion on the GPU from the dense form to GCOO is shown in Algorithm \ref{algo:gcooconv}.
\begin{algorithm}[!ht]
	\caption{convertToGCOOFormat}\label{algo:gcooconv}
	\textbf{Input: } $A, wA, hA, p$\\
	\textbf{Output: $values, cols, rows, gIdxes, nnzPerGroup$}
	\begin{algorithmic}[1]
		\small
		\State $nGroup = (hA + p - 1) / p$;
		\State Allocate memory for $gIdxes$ and $nnzPerGroup$ according to $nGroup$;
		\State Calculate $gIdxes$ and $nnzPerGroup$ and $nnz$ by scanning $\ma{A}$;
		\State Allocate memory for $values$, $cols$, and $rows$ according to $nnz$;
		\State Set values of $values$, $cols$ and $rows$ by scanning $\ma{A}$;
	\end{algorithmic}
\end{algorithm}

\subsection{GCOOSpDM: an efficient SpDM algorithm}
In the proposed algorithm GCOOSpDM, we focus on three factors that have major impact on the performance. 1) Data partition for the CUDA execution context \cite{nvidia2010programming}. 2) The coalesced memory access of global memory on the sparse matrix $\ma{A}$ and the two dense matrices $\ma{B}$ and $\ma{C}$. 3) When exploiting the faster memory on Nvidia GPUs with the shared memory, we guarantee that the access of the shared memory has no bank conflict. 4) After accessing a single element of the sparse matrix $\ma{B}$, we strive to calculate more results for $\ma{C}$, i.e., achieving higher operational intensity, so that we can achieve higher GFLOPS.

\begin{figure}[!h]
	\centering
	\includegraphics[width=\linewidth]{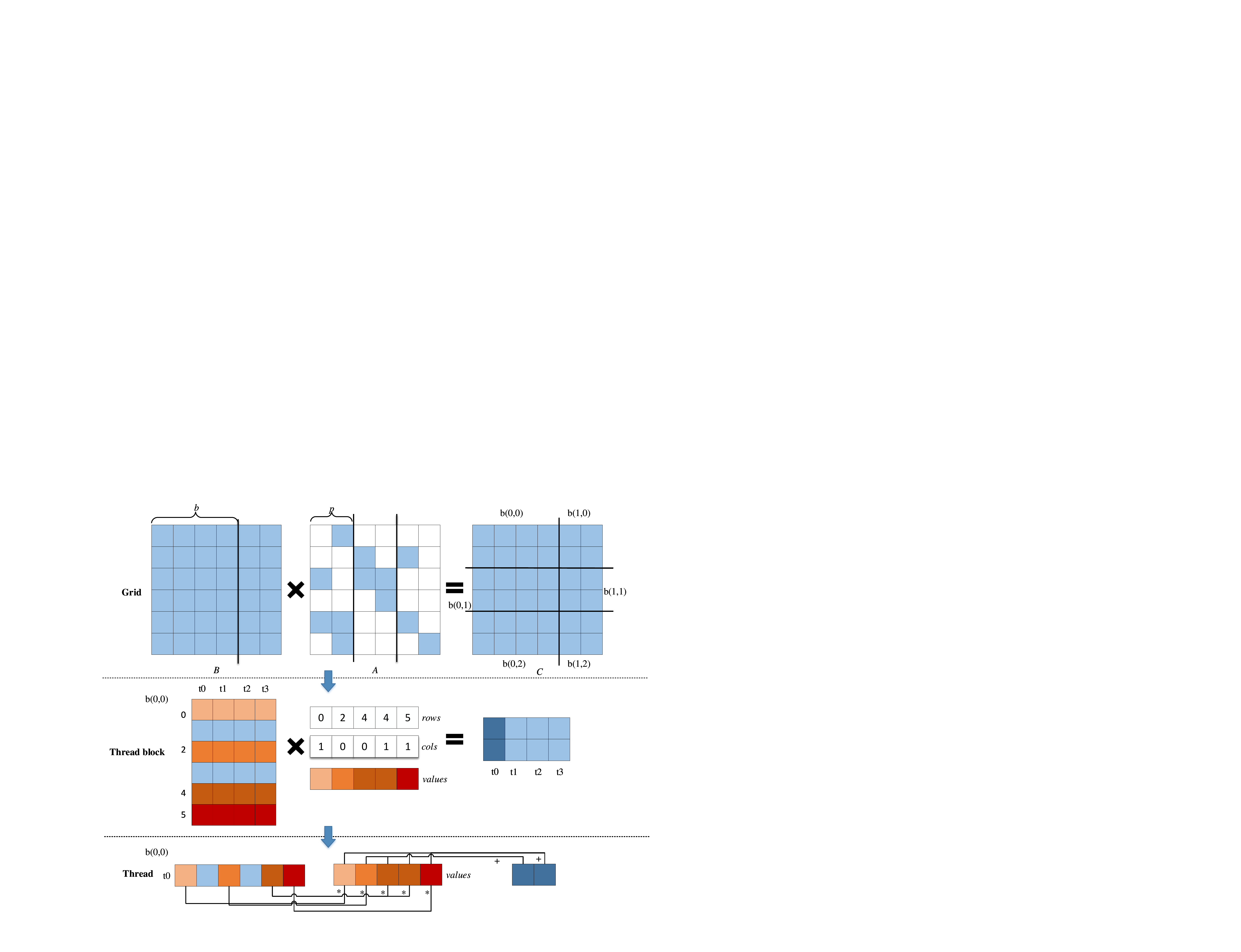}
	\caption{Partition of matrices. $\ma{A}$ is the sparse matrix, $\ma{B}$ is the dense matrix, and $\ma{C}$ is the result matrix.}
	\label{fig:multiplication}
\end{figure}

\textbf{Data partition of matrices}. In the context of CUDA, a \textit{thread} is the smallest execution unit of instructions. A group of threads forms a \textit{thread block}, which is executed in a stream multiprocessor (SM) of GPU. Multiple \textit{thread blocks} form a \textit{grid}, and some \textit{thread blocks} are executed in parallel on different SMs at one time. Let $b$ denote the size of a thread block. In our algorithm, each thread block calculates $b\times p$ elements of $\ma{C}$ separately, so a resulting $n\times n$ matrix requires $\lceil \frac{n}{b} \rceil\times \lceil \frac{n}{p} \rceil$ thread blocks. All threads in a thread block share a group of sparse data of $\ma{A}$, but each thread reads continuous columns $\ma{B}$ to do the operations of multiplication and addition to the continuous columns of $\ma{C}$. An example of data partition for $b=4, p=2$ and $n=6$ is shown in Fig. \ref{fig:multiplication}. In the grid, it has 6 thread blocks. Each thread block contains $b=4$ threads, and it calculates 8 elements of $\ma{C}$. Each thread calculates $p=2$ elements of $\ma{C}$.

\textbf{Coalesced memory access}. Three matrices including one sparse matrix $\ma{A}$ with the GCOO format and two dense arrays ($\ma{B}$ and $\ma{C}$) are needed to interactive with the global memory. Irregular global memory access would result in performance degradation on modern GPUs, so we should read the input matrices ($\ma{A}$ and $\ma{B}$) and write the output matrix $\ma{C}$ in a coalesced way. 

First, we consider the sparse matrix $\ma{A}$ stored with the GCOO format. Since each group in GCOO of $\ma{A}$ is assigned to one thread block, we just need to consider the block level access of one group of GCOO, i.e., a COO format that has $p$ columns. The number of floating point operations is determined by the number of nonzero elements of $\ma{A}$, so we scan COO to find the corresponding columns of $\ma{B}$. Due to the sparse property, COO could not have many elements, which means we can load COO to the shared memory such that all the threads can read the data fast. Therefore, the $b$ threads in one thread block read $b$ elements of COO from the global memory to the shared memory in a coalesced way. After $\ma{A}$ has been put into the shared memory, it is no need to re-read the elements of $\ma{A}$ from the global memory.

Second, the dense matrix of $\ma{B}$ should be read-aware. The matrix $\ma{B}$ only needs to be accessed when a $(col, row, a)$ of COO has been read from the shared memory, so every thread reads the same $(col, row, a)$, the corresponding column of $\ma{B}$ should be same while the rows should be different to keep all the threads busy and work balance. So threads $t_0, t_1, ..., t_{b-1}$ need to read $\ma{B}(row_0, col), \ma{B}(row_1, col), ..., \ma{B}(row_{b-1}, col)$ in the current block respectively. In order to support the coalesced memory read of $\ma{B}$, the row elements should be in the continuous memory. It is easy to do this because we can just transpose $\ma{B}$ or store $\ma{B}$ in a column-major matrix such that the above elements are in the continuous memory.

Finally, for the result matrix $\ma{C}$, we should only write the matrix once with the final result for each thread to achieve higher throughput. As discussed above, thread $t_i$ reads $(col, row, a)$ of $\ma{A}$, and multiplies with the elements indexed by $(row_i, col)$ in $\ma{B}$, so the write position of $\ma{C}$ should be $(row, row_i)$. As a result, $\ma{C}$ should also be column-major or transposed for the coalesced memory writing.

\textbf{None bank conflict access of the shared memory}. The shared memory used in our algorithm is only proper to the sparse matrix of $\ma{A}$ with the COO format (in one thread block). The kernel allocates a fixed size $b$ of shared memory, and the threads in one thread block read $b$ non-zero elements from $\ma{A}$ each time. Since all the threads in one thread block need to read all elements of $\ma{A}$ to calculate the corresponding columns of $\ma{C}$, all threads read the same element of $\ma{A}$. Therefore, the data in the shared memory can be accessed by all threads in a broadcast way \cite{nvidia2010programming}, which would not result in any bank conflict, and the broadcast access of the shared memory requires only a very small number of clock cycles to fetch the data.

\textbf{High computation-to-memory ratio}. Achieving a high operational intensity $r$ is very important to a high throughput. Regarding the multiplication and accumulation of each thread, each thread reads the shared memory of $\ma{A}$ to get $(col, row, a)$ (donated by $a_r$), and then multiplies $\ma{B}(row_i, col)$ (donated by $b_r$) of $\ma{B}$. In such scenario, we have two opportunities to have more calculations with $a_r$ and $b_r$ since they have been loaded into the registers. The first chance is to find other element of $\ma{B}$ to be multiplied with $a_r$, but the other element that can be multiplied with $a_r$ has been assigned to the other block, so this chance cannot be fulfilled. The second one is to find a next element of $\ma{A}$ who has the same column with the previous one while its row is different, i.e., $(col, row_1, a)$. Therefore, we can search the next $a_{r1}$ (since $\ma{A}$ has been loaded in the shared memory, the time cost of searching is low.) to reuse $b_r$. If such an $a_{r1}$ exists, then we can have $b$ times of multiplication and accumulation without an extra global memory (or L2 cache) access, which results in a higher $r$. For a uniform distributed sparse matrix with sparsity of $s$, there could be $(1-s)\times n$ non-zero elements in the same column. 

According to the above four criteria, we conclude the GCOOSpDM algorithm with the following three steps.

Step 1. Each thread block iteratively reads the COO values into the shared memory such that all threads in this thread block can read the COO values for their rows. We exactly know the columns that we need to calculate in the current thread block.

Step 2. The $t^{th}$ thread scans the COO items from the shared memory, and the item contains $row$, $col$ and $value$. According to $col$, the thread reads the element $B(t,col)$ of $\ma{B}$, and then performs the multiplication of $value \times B(t, col)$, whose result is added to the local variable $c_{t,col}$. I.e., $c_{t,col}+=value \times B(t, col)$. 

Step 3. Since the current group of data is stored as the COO format, for the current element $(row, col, value)$, its next element should have the same $col$ index if that column has more than one element. So we continue scanning the shared memory to check if there are elements that have the same $col$ such that we can reuse the element of $B(t, col)$. 

The visualization of the algorithm executed with the CUDA programming model is shown in Fig. \ref{fig:multiplication}. On the grid level, there are 6 thread blocks, and each thread block calculates the results of sub-matrix with size of $b\times p$ from $b$ rows of $\ma{B}$, and $p$ columns (i.e., one group in GCOO) of $\ma{A}$. On the thread block level, the GCOO data of sparse matrix are loaded into faster memory once (the shared memory) which is shared among all the threads in the thread block. On the thread level, each thread independently takes charge of computing $p$ elements of $\ma{C}$, say the thread scans the shared memory to read $row$, $col$ and $value$, and then reads the values in column $row$ of $\ma{B}$, which are multiplied by $value$ separately, and each result is accumulated to column $col$ of $\ma{C}$. The algorithm of GCOOSpDM is shown in Algorithm \ref{algo:gcoospdm}.

In Algorithm \ref{algo:gcoospdm}, we first (line 1-10) initialize some local variables including the thread level indexes of output and COO for the current thread block. Then we iteratively scan a block of COO in the for-loop of line 11, and at each iteration, a thread block of COO values are loaded into the shared memory (line 12-15). After that each value of COO in the shared memory is read by all the threads in one thread block, and the corresponding value $b$ in $\ma{B}$ is also read to calculate the result (line 21-26). Instead of continuing the above step, we keep the value of $b$ in the register, and scan the shared COO to check whether we can reuse $b$ so that less memory operations are required (line 28-36). By this way, we can achieve higher operational intensity, i.e., $b$ is reused to do more floating point calculations. At the end, the local results of each thread are written back to $\ma{C}$ that is stored in the global memory with corresponding indexes (line 38-39). Note that both reading of matrix $\ma{A}$ and matrix $\ma{B}$ from the global memory is in a coalescent way, the result writing to matrix $\ma{C}$ is also coalescent. In term of access of the shared memory, it broadcast the data to all the threads in a warp with a small number of cycles.

\begin{algorithm}[!ht]
	\caption{GCOOSpDM}\label{algo:gcoospdm}
	\textbf{Input: } $values, cols, rows, gIdxes, nnzPerGroup, wA, hA, \\B, wB, hB, C$\\
	\textbf{Output: $C$}
	\begin{algorithmic}[1]
		\small
		\State $Cj = blockIdx.y*b+threadIdx.x$;
		\State $Ci0 = blockIdx.x*p$;
		\State Initial local temporary results $c[0...p]$;
		\State Set number of non-zero elements of current group: $nnz$;
		\State // Set the current group of COO
		\State $vals=values+gIdxes[blockIdx.x]$;
		\State $cols=cols+gIdxes[blockIdx.x]$;
		\State $rows=rows+gIdxes[blockIdx.x]$;
		\State $iter=(b+nnz-1)/b$;
		\State $extra = nnz \& (b - 1)$; 
		\For{$i=0 \rightarrow iter$}
			\State $cooOffset=i*b$;
			\State $sVals[threadIdx.x]=vals[cooOffset]$;
			\State $sCols[threadIdx.x]=cols[cooOffset]$;
			\State $sRows[threadIdx.x]=rows[cooOffset]$;
			\State $cnnz=\text{max}(extra, b)$;
			\State $\_\_$syncthreads();
			\If{$Cj<wB$} // Not exceed the boundary
				\State $k=1$;
				\For{$j=0\rightarrow cnnz, step=k$}
					\State $col = sCols[j]$;
					\State $row = sRows[j]$;
					\State $av=sVals[j]$;
					\State $bv=B[col*wB+Cj]$; // Registered.
					\State $outIdx=row\&(p-1)$;
					\State $c[outIdx]+=av*bv$;
					\State $k=1$;
					\While{$j+k<cnnz$} // Search $A$ to reuse $bv$
						\State $newCol=sCols[j+k]$;
						\If{$newCol \neq col$}
							\State break;
						\EndIf
						\State $av=sVals[k+j]$;
						\State $row = sRows[k+j]$;
						\State $outIdx=row\&(CPG-1)$;
						\State $c[outIdx]+=av*bv$;
						\State $k+=1$;
					\EndWhile
				\EndFor
			\EndIf
			\State $\_\_$syncthreads();
		\EndFor
	\For{$i=0 \rightarrow p$} // Write results to the global memory
	\State $C[Cj+(Ci0+i)*wB]=c[i]$;
	\EndFor
	\end{algorithmic}
\end{algorithm}

\section{Evaluation and Analysis}\label{section:evaluation}
To show the effectiveness of our proposed algorithm, we do varies of experiments across three Nvidia GPU cards (i.e., GTX 980, GTX Titan X Pascal and Tesla P100) using two kinds of data. The first one is the public sparse matrix dataset \cite{davis2011university} which has different patterns of matrices, and the second one is randomly generated matrices whose zero-valued elements have a uniform distribution.\footnote{Codes of GCOOSpDM and scripts of performance evaluation can be found in \url{https://github.com/hclhkbu/gcoospdm}. And the raw data of our experimental results can be found in: \url{https://github.com/hclhkbu/gcoospdm/tree/master/results}.} The characteristics of tested GPUs are shown in Table \ref{table:gpus}. And the software installed is CUDA-8.0.
\begin{table}[!ht]
	\centering
	\caption{Characteristics of tested GPUs.}
	\label{table:gpus}
	\begin{tabular}{|l|l|l|l|}
		\hline
		Model& 	GTX980 &TitanX  & P100 \\\hline
		\hline
		SMs $\times$ cores per SM & 16$\times$128 &  28$\times$128 & 56$\times$64 \\\hline
		Peak TFLOPS &4.981  	& 10.97	& 9.5 \\\hline
		Memory Bandwidth (GB/s) &224   & 433	& 732 \\\hline
	\end{tabular}
\end{table}

\subsection{Results on public sparse matrices}
\begin{figure}[!h]
	\centering
	\subfigure[GTX 980]
	{
		\includegraphics[width=0.32\linewidth]{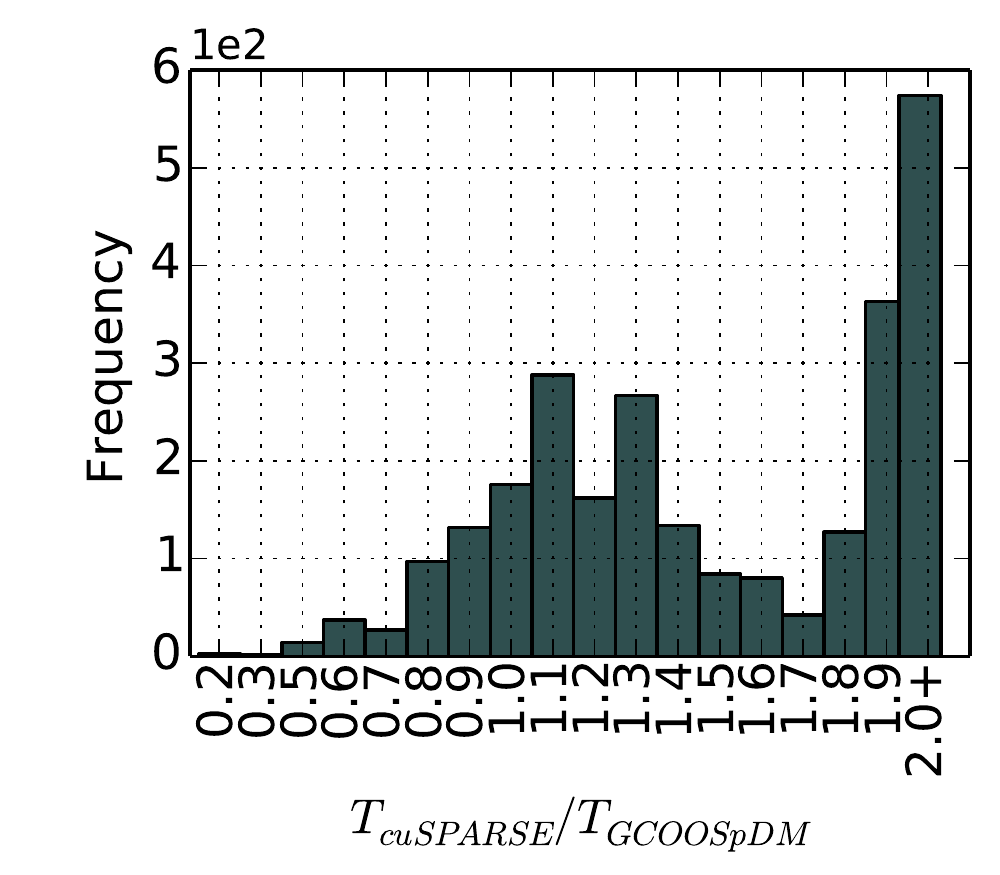}
	}\hspace{-3.5mm}
	\subfigure[Titan X Pascal]
	{
		\includegraphics[width=0.32\linewidth]{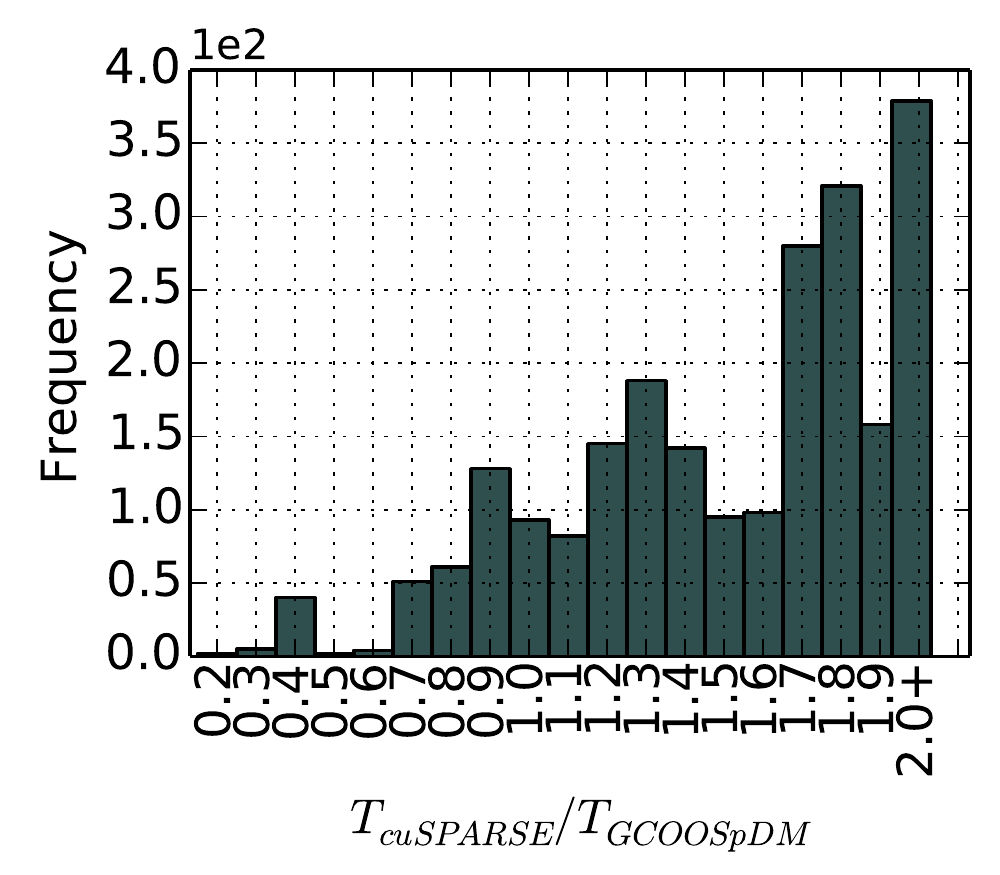}
	}\hspace{-3.5mm}
	\subfigure[Tesla P100]
	{
		\includegraphics[width=0.32\linewidth]{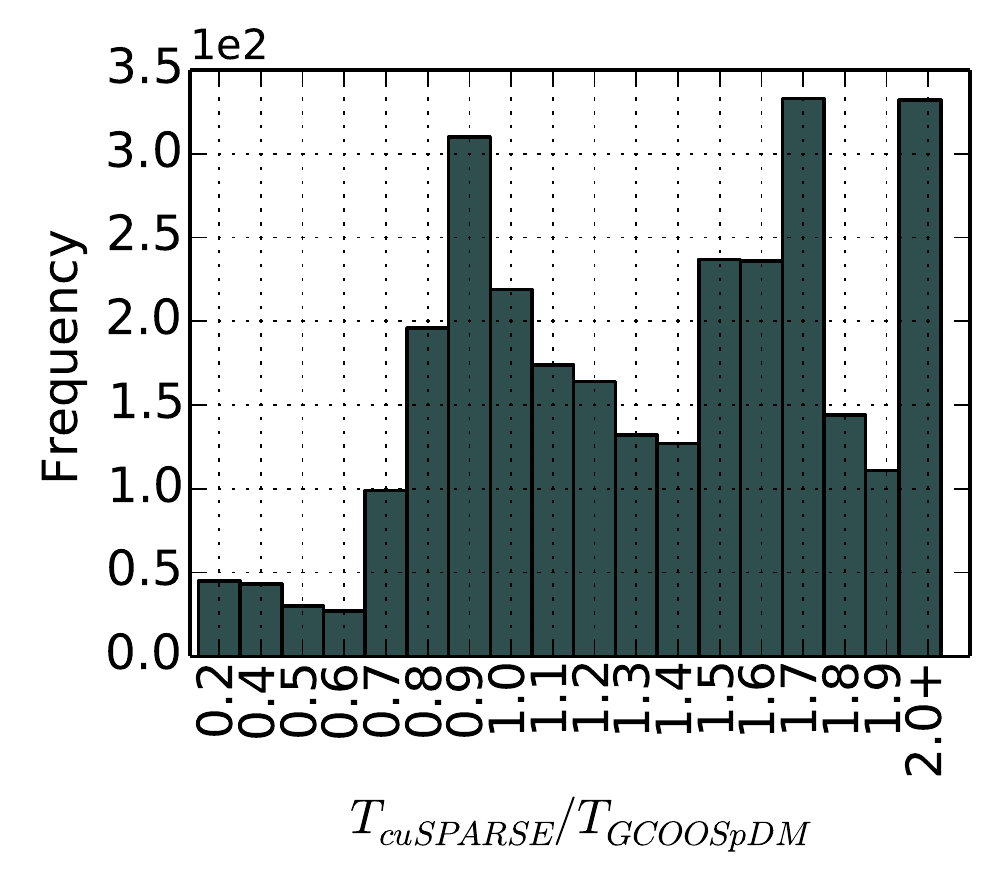}
	}
	\caption{The performance comparison with the frequency of the time ratio between cuSPARSE and GCOOSpDM with the public dataset on three GPUs. The last value (i.e., 2.0+) of x-axis means that $T_{cuSPARSE}/T_{GCOOSpDM} \ge 2.0$.}
	\label{fig:realdata}
\end{figure}

We use the public sparse matrices in \cite{davis2011university}. Since we only consider the schemes of square matrices, we pick up all the square matrices in the dataset to evaluate the performances of GCOOSpDM and cuSPARSE. The chosen dataset contains 2694 matrices, whose sparsity is in the range of $[0.98, 0.999999]$, and their dimensions are in the range of $[64, 36720]$. The performance comparison between GCOOSpDM and cuSPARSE is shown in Fig. \ref{fig:realdata}, where $T_{algorithm}$ is used to denote the execution time of $algorithm$. We first compare the overall performance of our algorithm with cuSPARSE on the 2694 matrices, and we then choose 14 types of matrices from varies of applications to compare the performance of the algorithms. 

\textbf{Overall performance}. In the 2694 tested matrices, there are about $78\%$ matrices that GCOOSpDM outperforms cuSPARSE on the P100 GPU, and there are more than $90\%$ matrices that GCOOSpDM achieves better performance than cuSPARSE on both GTX980 and TitanX. The average speedups are $1.66\times$, $1.7\times$ and $1.68\times$ on GTX980, TitanX and P100 respectively. Moreover, the maximum speedups of GCOOSpDM are $4.5\times$, $6.3\times$ and $4.2\times$ on GTX980, TitanX and P100 GPUs respectively. By contrast, on the $22\%$ matrices that cuSPARSE is better than GCOOSpDM on the P100 GPU, cuSPARSE only outperforms GCOOSpDM about $1.2\times$ on average. On GTX 980 and Titan X GPUs, there are about $10\%$ cuSPARSE outperforming GCOOSpDM about $1.14\times$. cuSPARSE performs better on the P100 GPU than GTX 980 and TitanX GPUs mainly because the P100 GPU has a much higher memory bandwidth than the other two GPUs as shown in Table \ref{table:gpus}.
\begin{table}[!ht]
	\centering
	\caption{Details of selected sparse matrices.}
	\label{table:selectedmatrx}
	\begin{tabular}{|l|l|l|l|}
		\hline
		Matrix & $n$ & Sparsity & Related Problem\\\hline
		\hline
		nemeth11&9506&2.31e-03&Quantum Chemistry \\\hline
		human\_gene1&22283&2.49e-02&Undirected Weighted Graph\\\hline
		Lederberg&8843&5.32e-04&Directed Multigraph\\\hline
		m3plates&11107&5.38e-05&Acoustics \\\hline
		aug3dcqp&35543&6.16e-05&2D/3D \\\hline
		Trefethen\_20000b&19999&7.18e-04&Combinatorial \\\hline
		ex37&3565&5.32e-03&Computational Fluid\\\hline
		g7jac020sc&5850&1.33e-03&Economic \\\hline
		LF10000&19998&1.50e-04&Model Reduction \\\hline
		epb2&25228&2.75e-04&Thermal \\\hline
		plbuckle&1282&9.71e-03&Structural \\\hline 
		wang3&26064&2.61e-04&Semiconductor Device \\\hline
		fpga\_dcop\_01&1220&3.96e-03&Circuit Simulation \\\hline 
		viscoplastic2\_C\_1&32769&3.55e-04&Materials \\\hline
	\end{tabular}
\end{table}

\begin{figure}[!ht]
	\centering
	\includegraphics[width=0.9\linewidth]{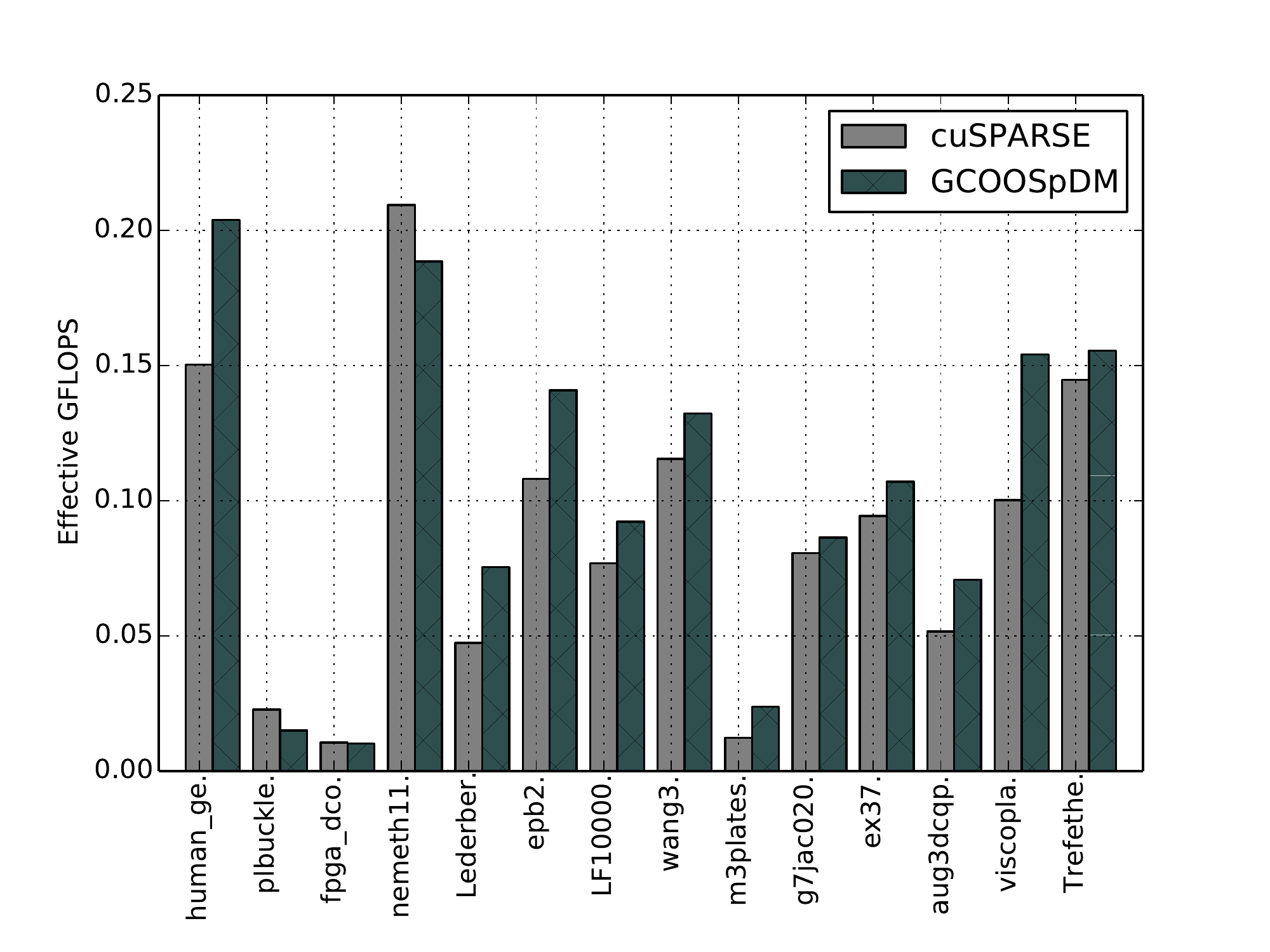}
	\caption{The performance comparison of selected matrices on a Tesla P100 GPU. (The higher the better.)}
	\label{fig:realbar}
\end{figure}

\textbf{14 types of matrices}. It can be seen that GCOOSpDM does not always outperform cuSPARSE. To further understand the main reasons, we select 14 types of matrices that have different structures and non-zero patterns from a range of areas to analyze their performance differences. The details of the selected matrices are shown in Table \ref{table:selectedmatrx}. To normalize the algorithm performances, we use effective GFLOPS to measure the algorithms as the following Equation 
\begin{equation}\label{equ:perf}
P_{algorithm}=\frac{2\times n^3\times (1-s)}{T_{algorithm}}.
\end{equation}
The performance comparison is shown in Fig. \ref{fig:realbar}. On three matrices (``nemeth11'', ``plbuckle'' and ``fpga\_dcop\_01''), GCOOSpDM is worse than cuSPARSE due to the non-zero distribution of the matrices. On these three matrices, the non-zero elements are mainly located on the diagonal of the matrices, such that there is little opportunity to reuse the pre-fetched value of $bv$ (i.e., line 30 will intermediately hold and no further calculations for current $bv$), but it still spends extra overheads to search $\ma{A}$.

\subsection{Random sparse matrices}
We randomly generate square matrices whose dimension are in the range of $[400, 14500]$ with a step size of $100$. For each size of a matrix, we generate the elements with the sparsity in two ranges (i.e, $[0.8,0.995]$ at a $0.005$ step and $[0.995, 0.9995]$ at a $0.0005$ step). In total, there are $6968$ matrices with uniformly distributed non-zero elements for evaluation. 

\begin{figure}[!h]
	\centering
	\subfigure[GTX 980]
	{
		\includegraphics[width=0.32\linewidth]{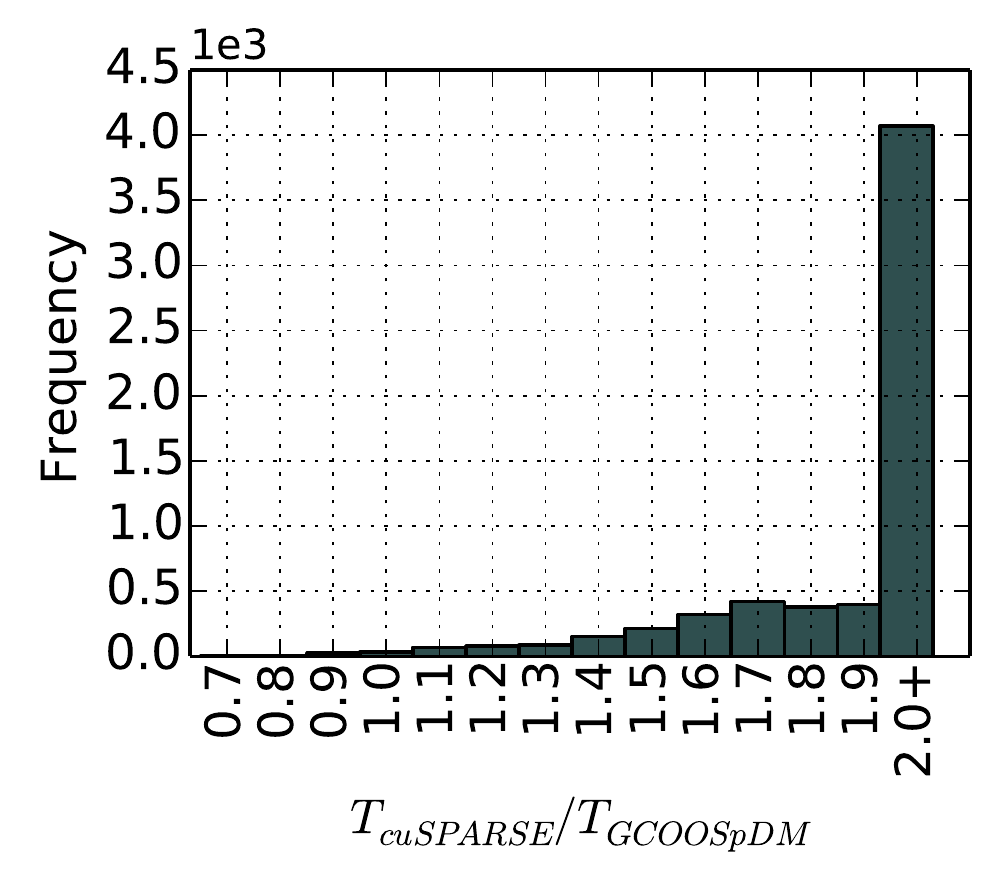}
	}\hspace{-3.5mm}
	\subfigure[Titan X Pascal]
	{
		\includegraphics[width=0.32\linewidth]{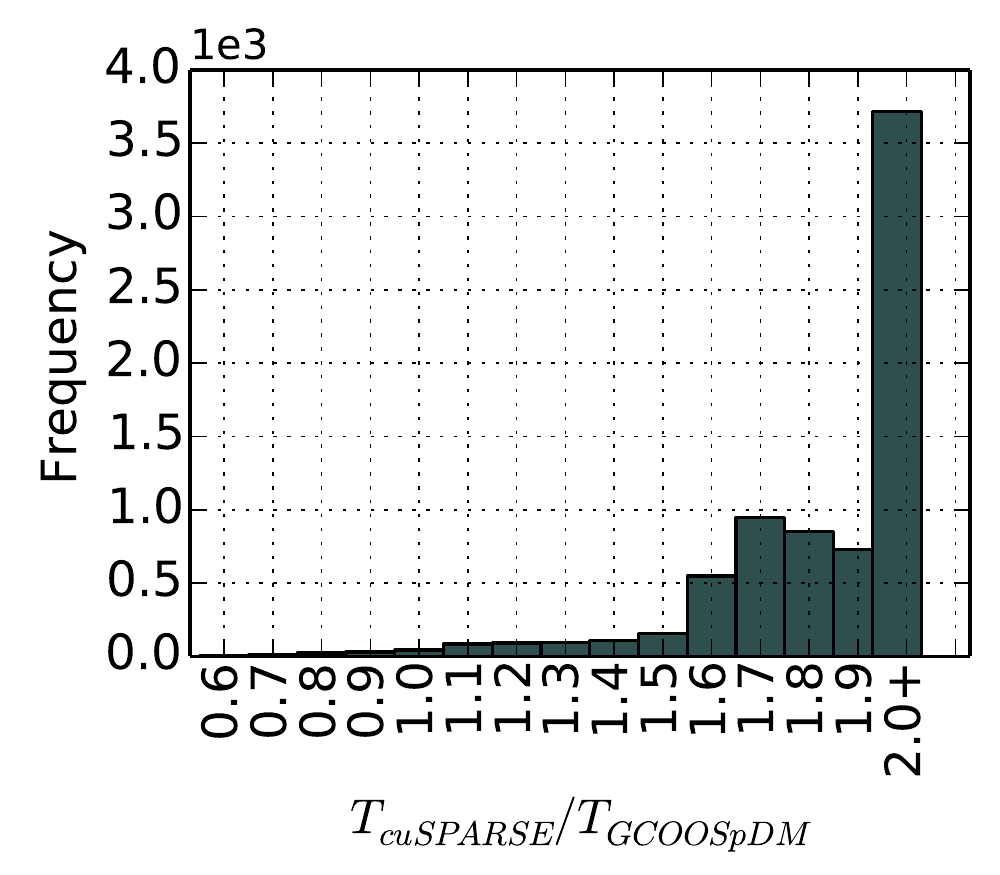}
	}\hspace{-3.5mm}
	\subfigure[Tesla P100]
	{
		\includegraphics[width=0.32\linewidth]{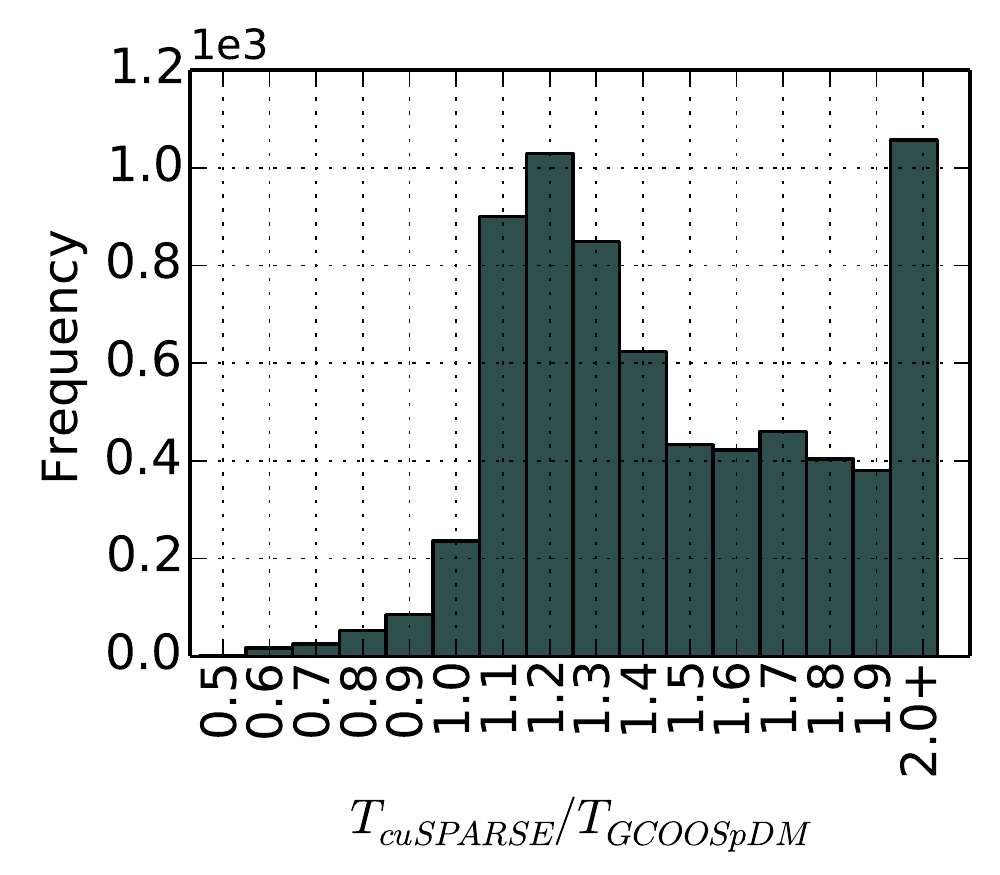}
	}
	\caption{The performance comparison with the frequency of the time ratio between cuSPARSE and GCOOSpDM with the random generated sparse matrices on three GPUs. The last value (i.e., 2.0+) of x-axis means that $T_{cuSPARSE}/T_{GCOOSpDM} \ge 2.0$.}
	\label{fig:sythperf}
\end{figure}

\textbf{Overall performance}. The performance comparison between GCOOSpDM and cuSPARSE using the randomly generated matrices is shown in Fig. \ref{fig:sythperf}. Our GCOOSpDM algorithm outperforms cuSPARSE in $99.51\%$, $99.23\%$ and $97.37\%$ matrices on GTX980, TitanX and P100 GPUs respectively, and the average speedups are $2.13\times$, $2\times$ and $1.57\times$ respectively. Particularly, the maximum speedups on the three GPUs are $4.7\times$, $6.5\times$ and $8.1\times$ respectively. On the cases that cuSPARSE is better GCOOSpDM, they only occupy a very small proportion (less than $3\%$), and the average performance ratio is only around $1.17$, which indicates very close performance on less than $3\%$ cases. 
\begin{figure}[!h]
	\centering
	\subfigure[$n=4000$]
	{
		\includegraphics[width=0.48\linewidth]{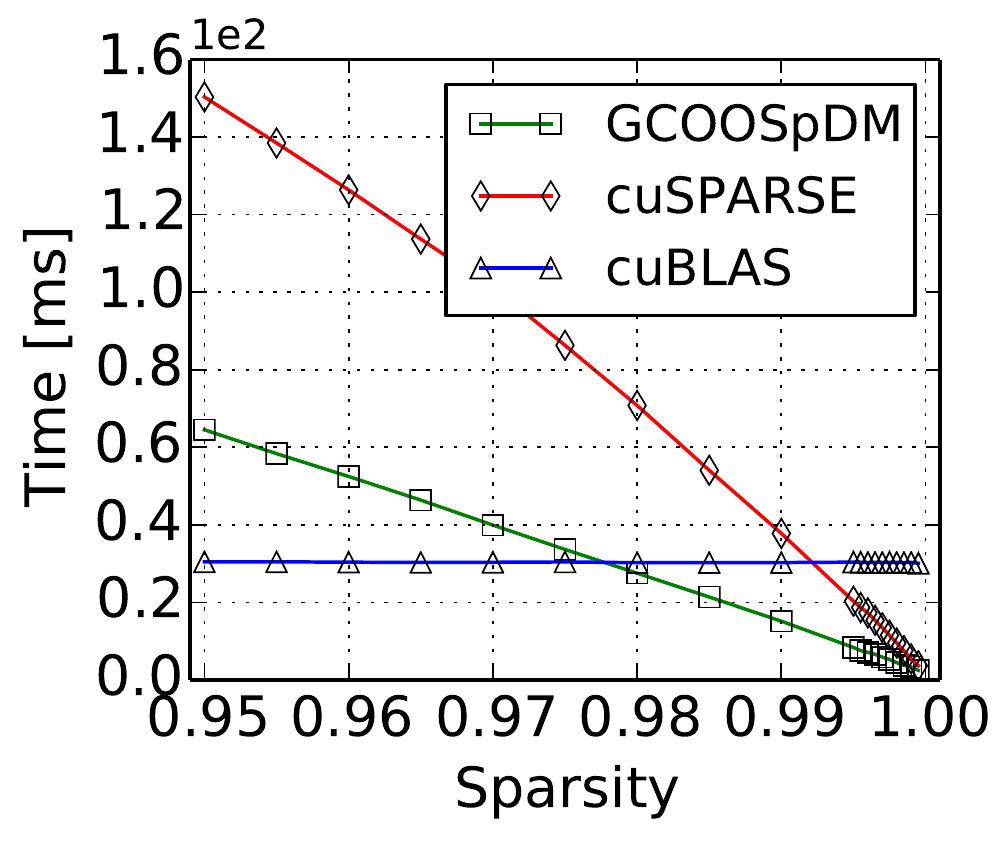}
	}\hspace{-3mm}
	\subfigure[$n=14000$]
	{
		\includegraphics[width=0.48\linewidth]{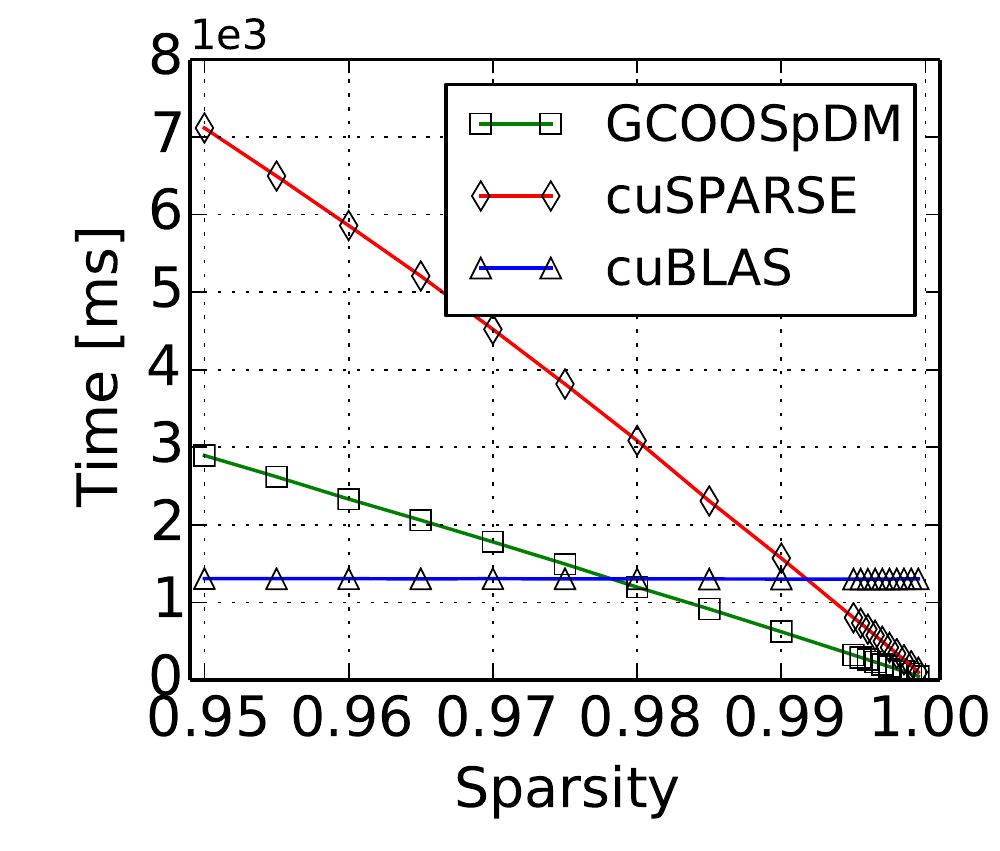}
	}
	\caption{Performance vs. sparsity on the GTX980 GPU. The lower the better.}
	\label{fig:perfvssparsity980}
\end{figure}
\begin{figure}[!h]
	\centering
	\subfigure[$n=4000$]
	{
		\includegraphics[width=0.48\linewidth]{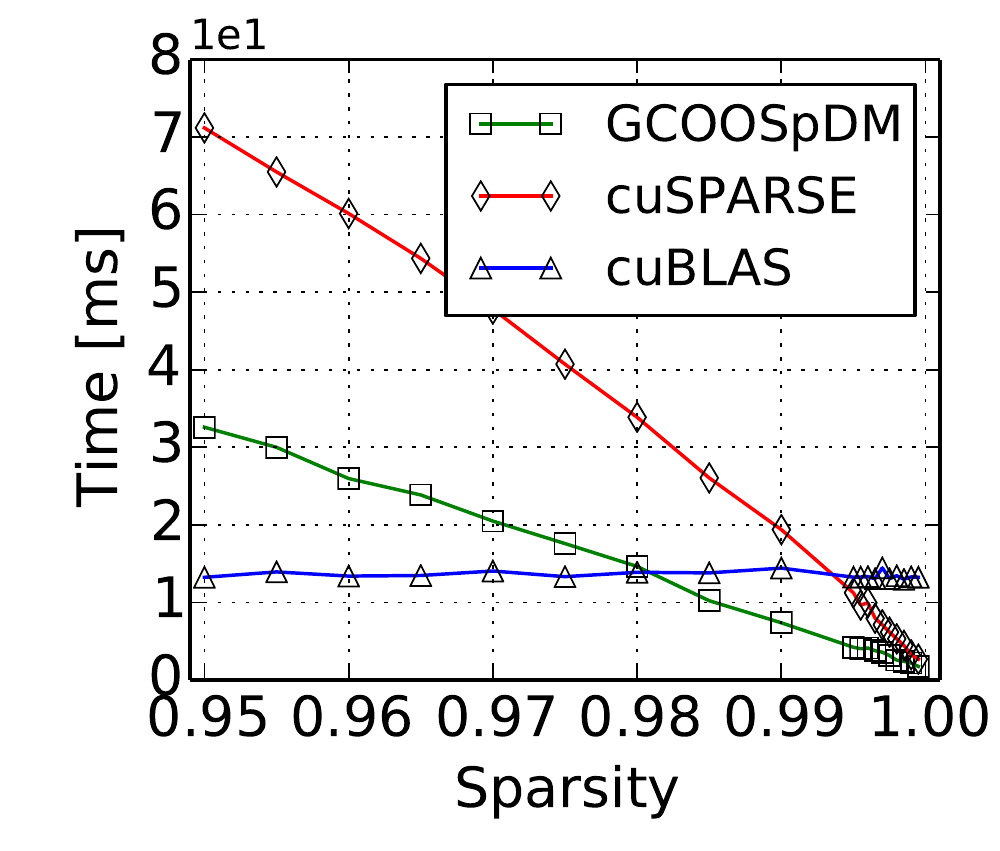}
	}\hspace{-3mm}
	\subfigure[$n=14000$]
	{
		\includegraphics[width=0.48\linewidth]{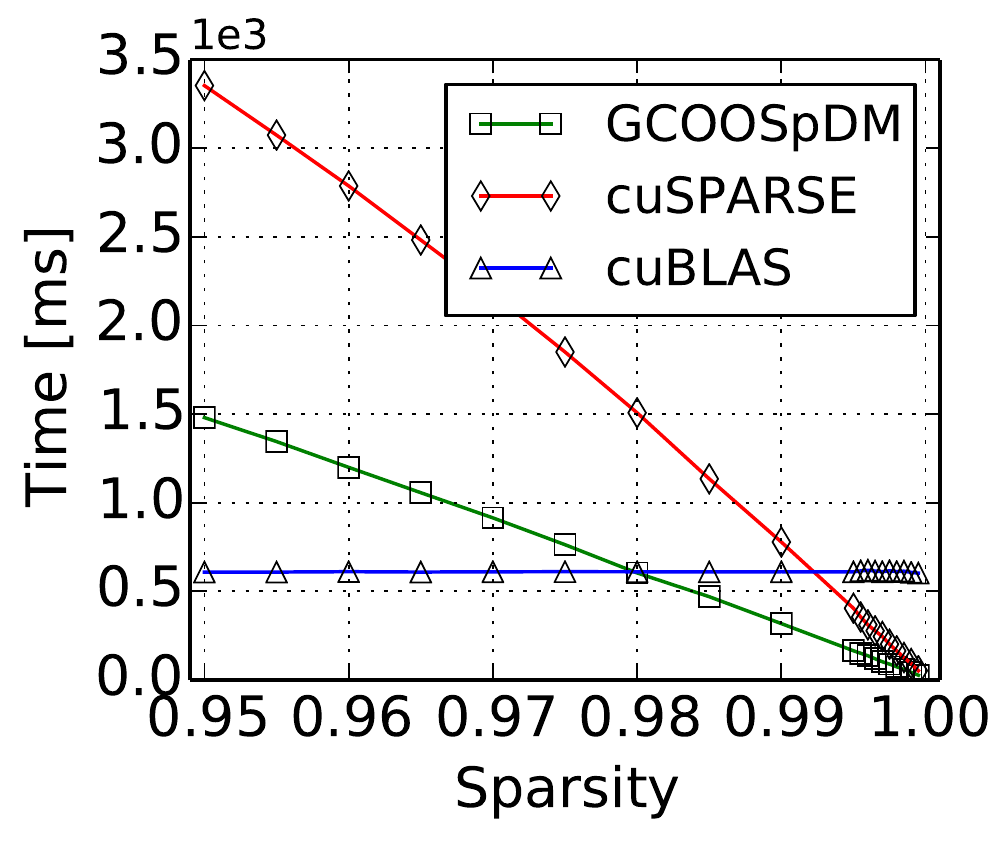}
	}
	\caption{Performance vs. sparsity on the TitanX GPU. The lower the better.}
	\label{fig:perfvssparsitytitanx}
\end{figure}
\begin{figure}[!h]
	\centering
	\subfigure[$n=4000$]
	{
		\includegraphics[width=0.48\linewidth]{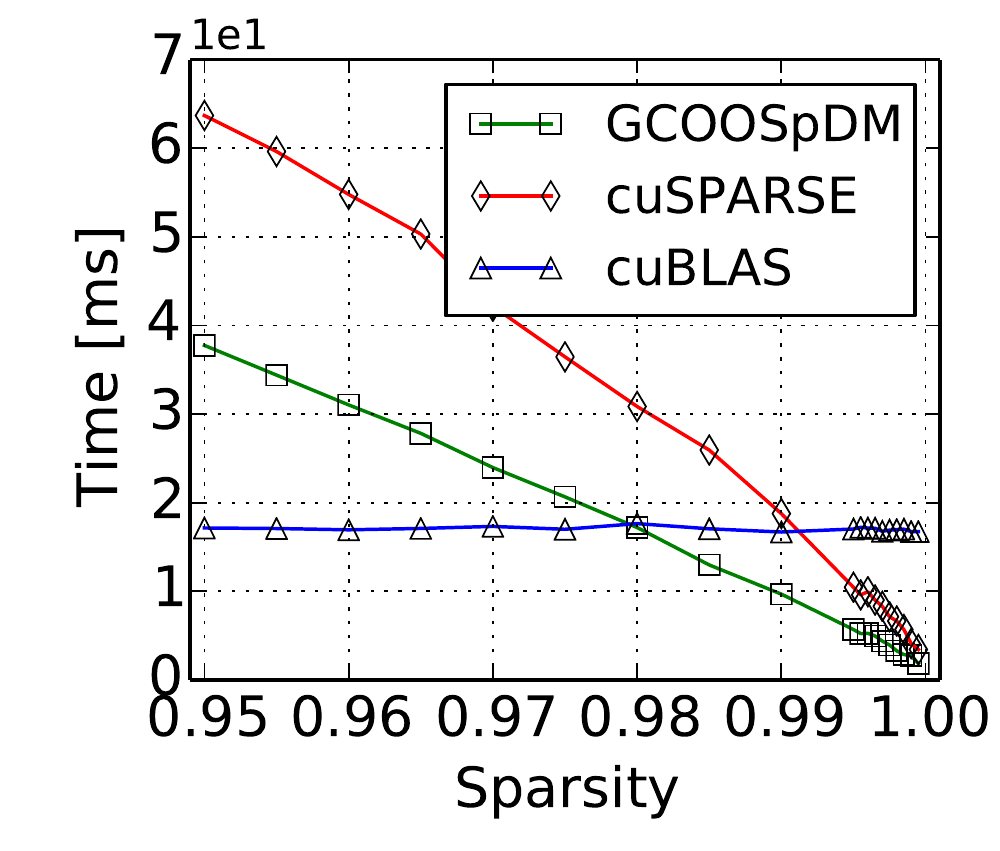}
	}\hspace{-3mm}
	\subfigure[$n=14000$]
	{
		\includegraphics[width=0.48\linewidth]{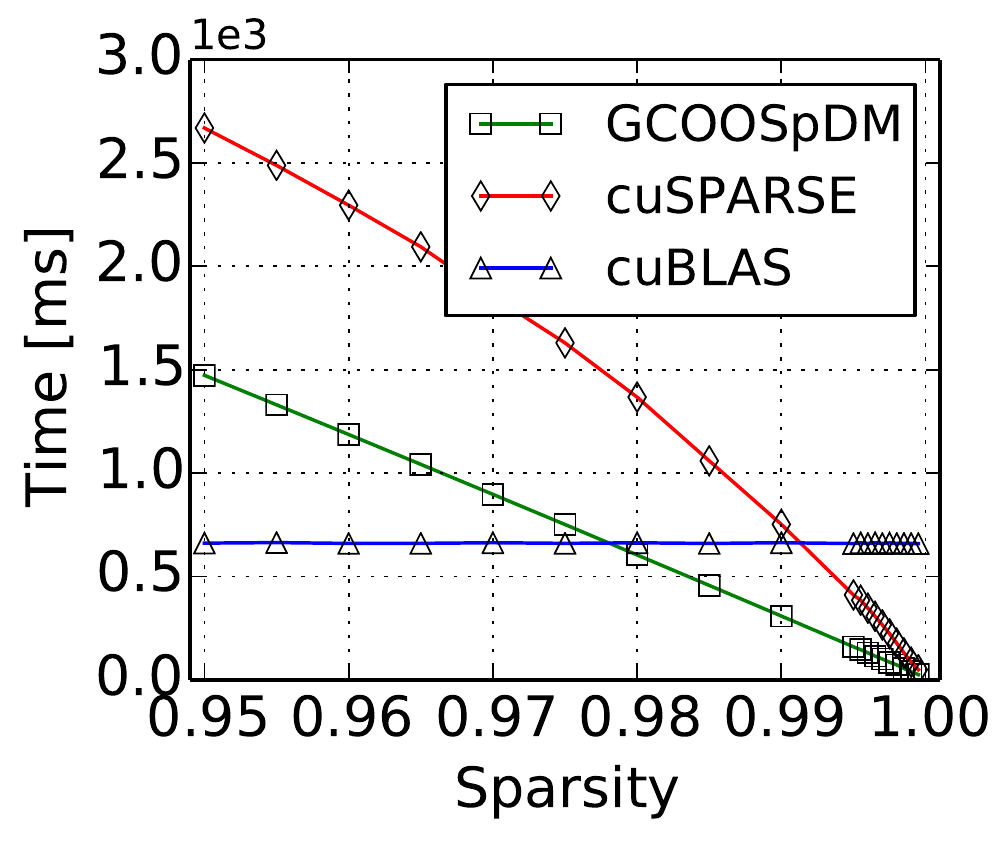}
	}
	\caption{Performance vs. sparsity on the P100 GPU. The lower the better.}
	\label{fig:perfvssparsityp100}
\end{figure}

\begin{figure}[!h]
	\centering
	\subfigure[$s=0.98$]
	{
		\includegraphics[width=0.48\linewidth]{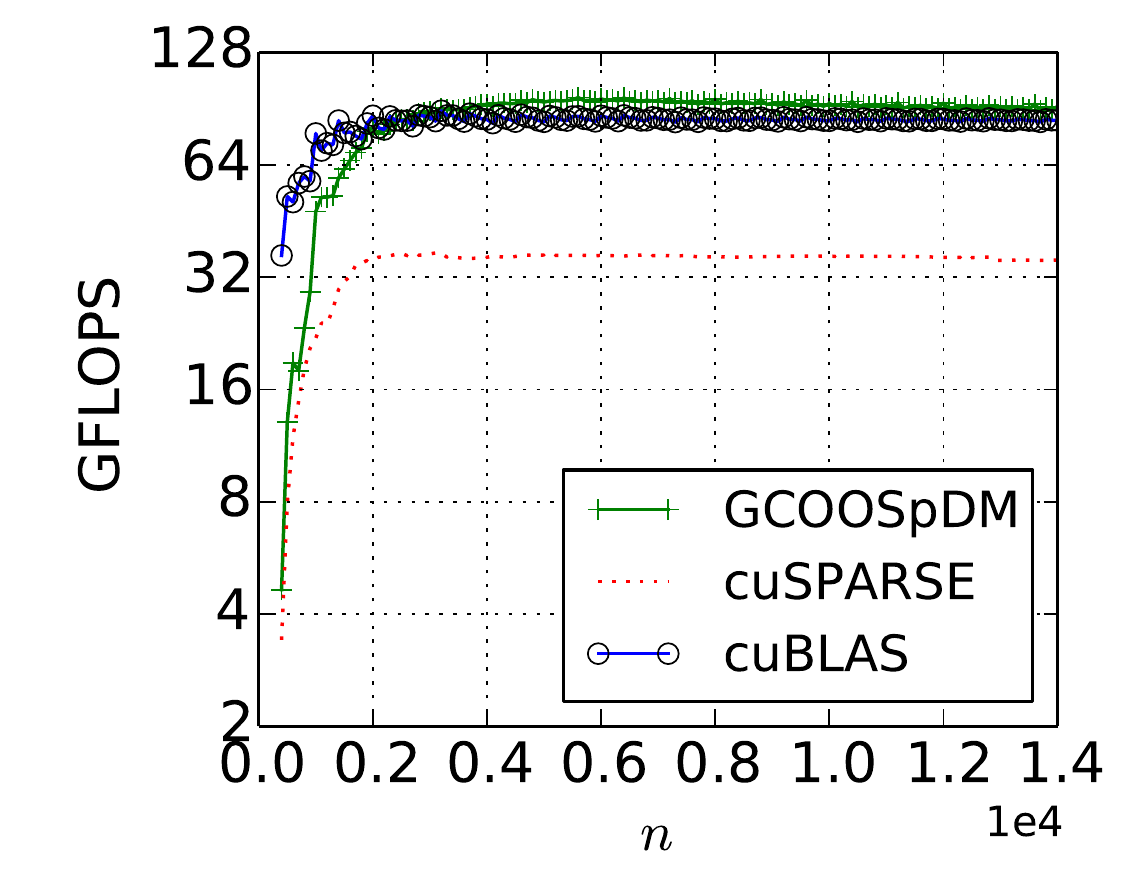}
	}\hspace{-3mm}
	\subfigure[$s=0.995$]
	{
		\includegraphics[width=0.48\linewidth]{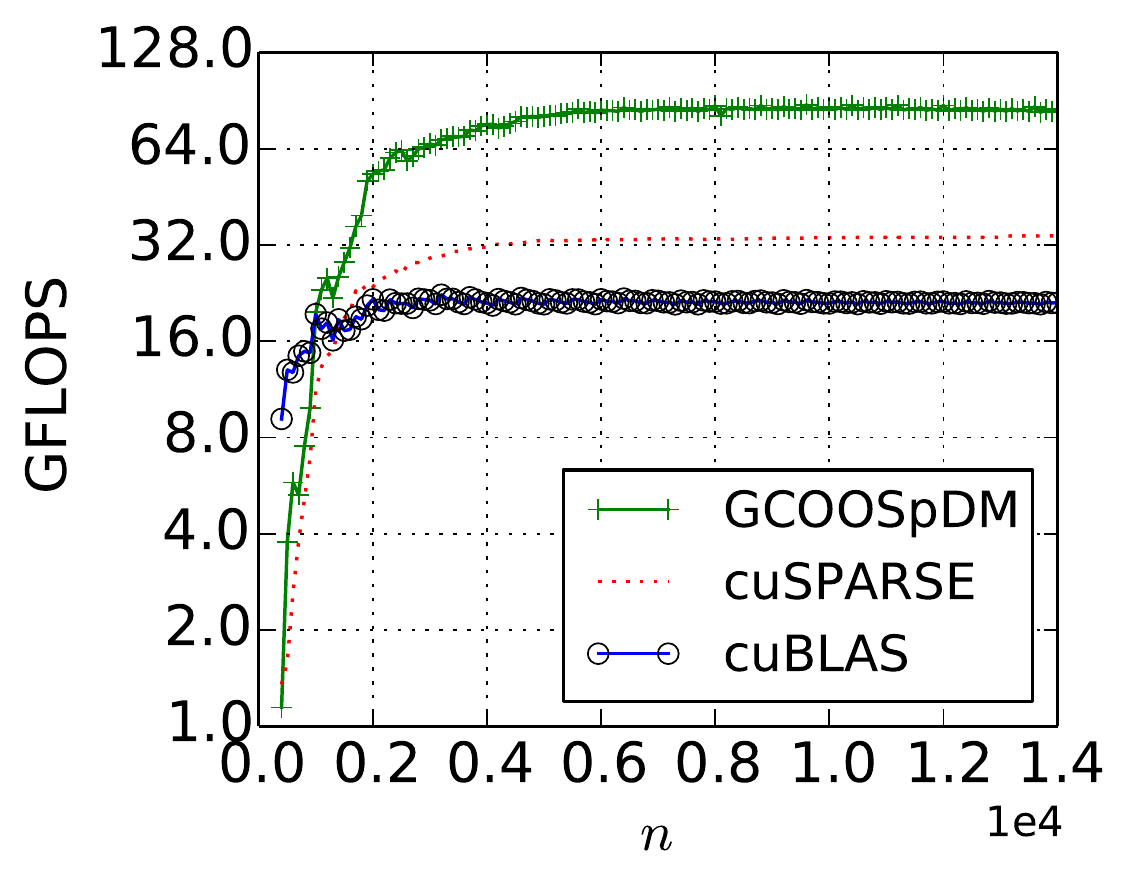}
	}
	\caption{Performance vs. dimension on GTX980. The higher the better.}
	\label{fig:perfwithspar980}
\end{figure}

\begin{figure}[!h]
	\centering
	\subfigure[$s=0.98$]
	{
		\includegraphics[width=0.48\linewidth]{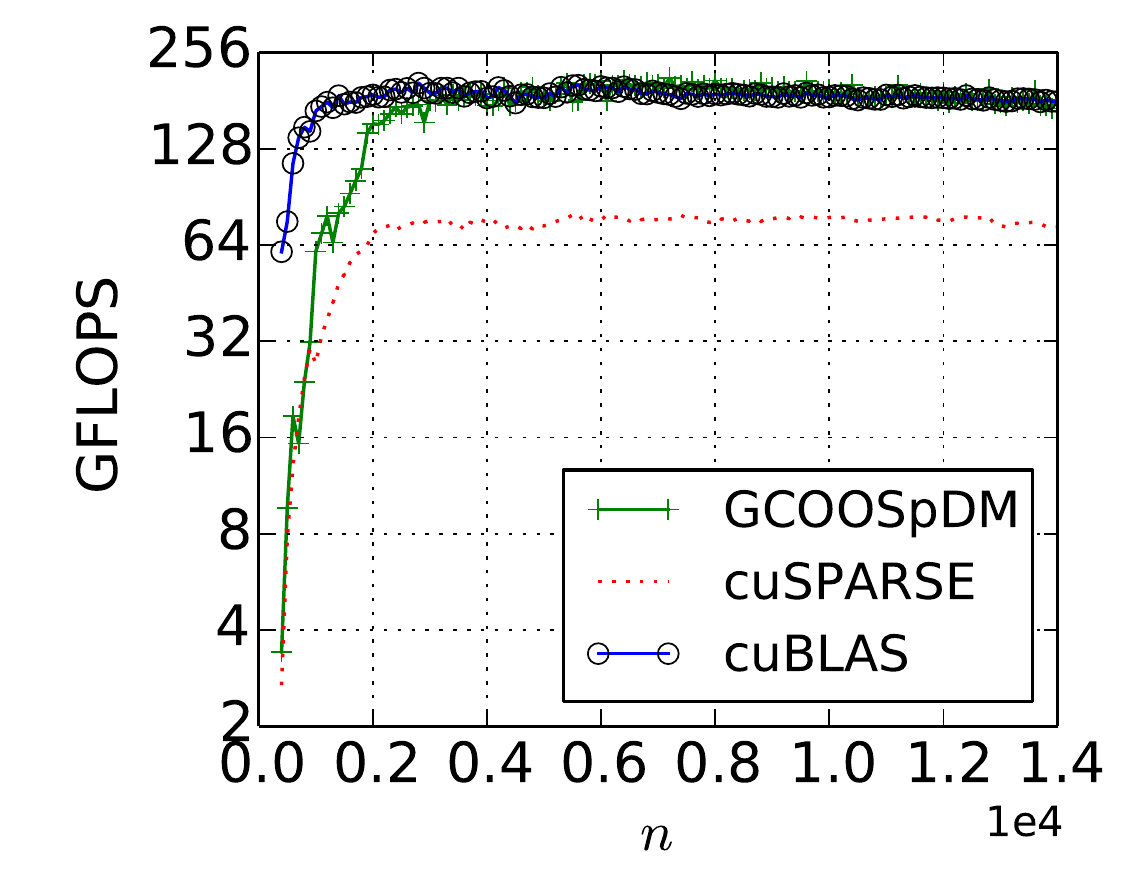}
	}\hspace{-3mm}
	\subfigure[$s=0.995$]
	{
		\includegraphics[width=0.48\linewidth]{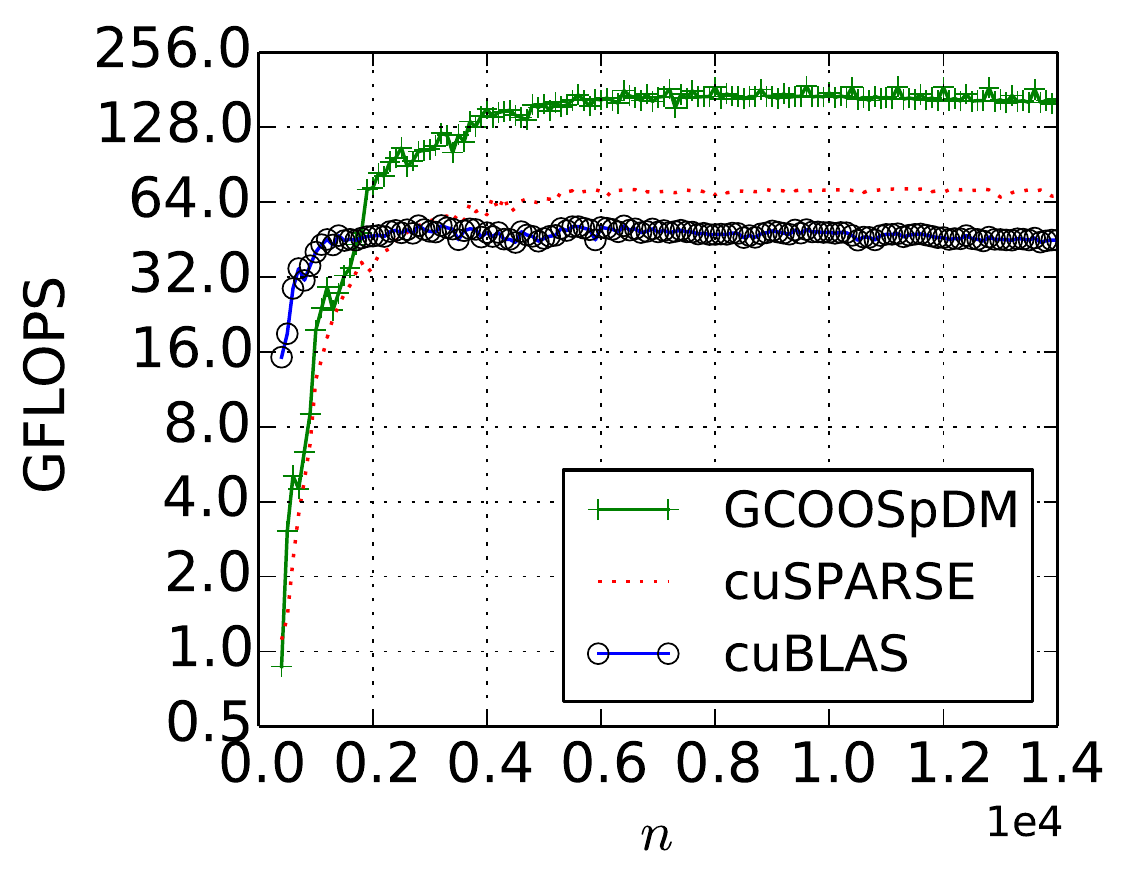}
	}
	\caption{Performance vs. dimension on TitanX. The higher the better.}
	\label{fig:perfwithspartitanx}
\end{figure}

\begin{figure}[!h]
	\centering
	\subfigure[$s=0.98$]
	{
		\includegraphics[width=0.48\linewidth]{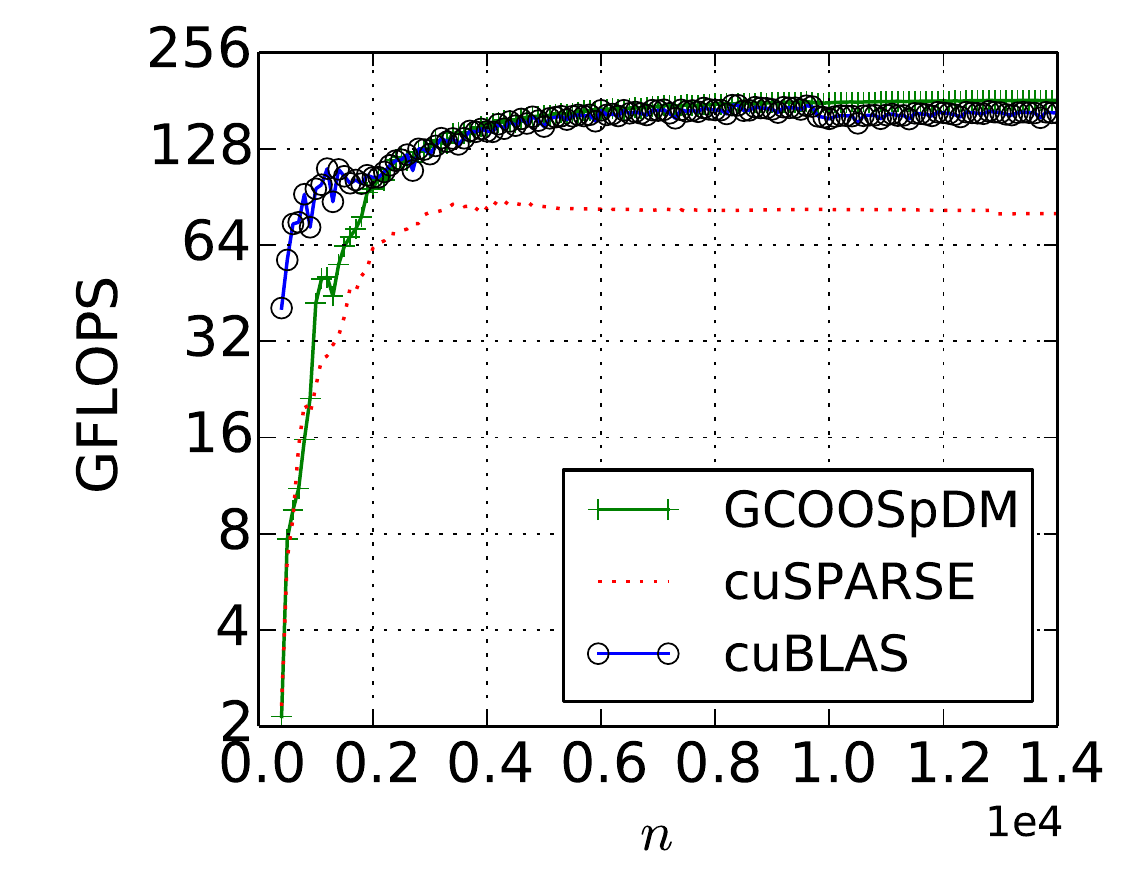}
	}\hspace{-3mm}
	\subfigure[$s=0.995$]
	{
		\includegraphics[width=0.48\linewidth]{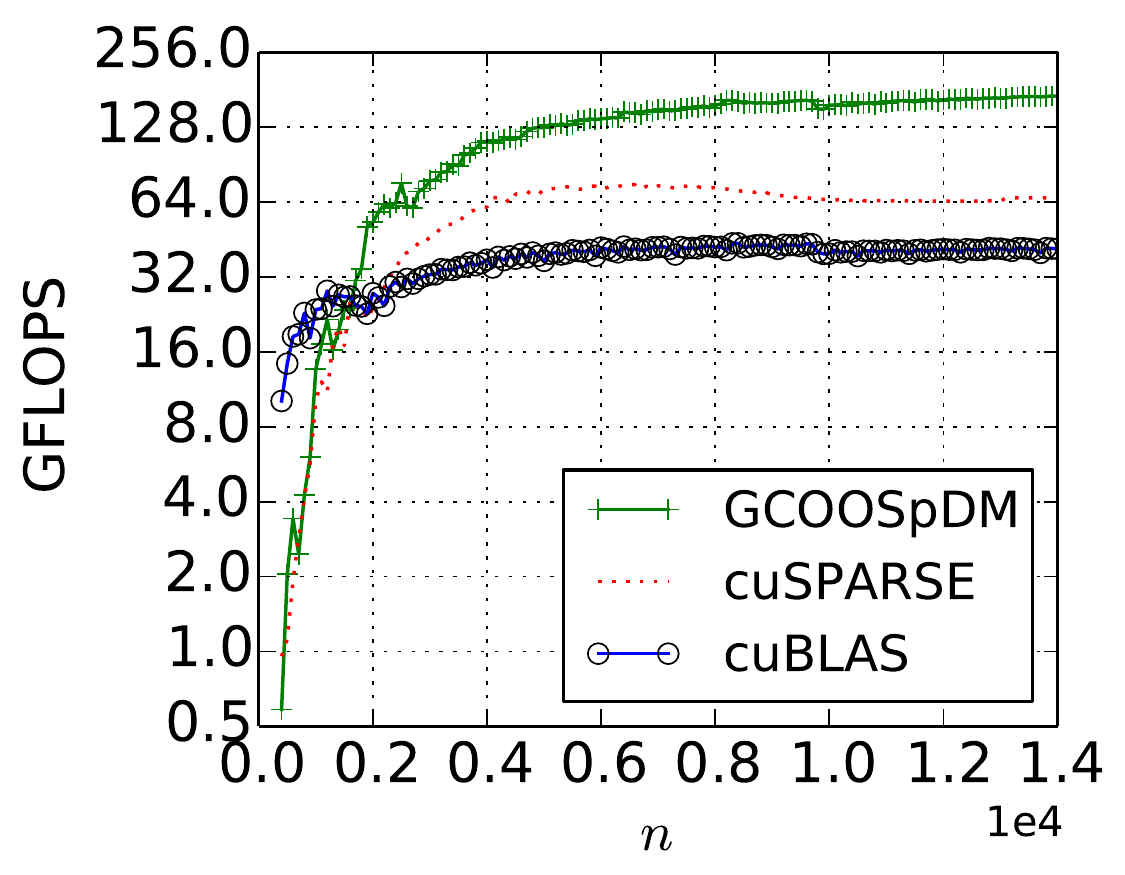}
	}
	\caption{Performance vs. dimension on P100. The higher the better.}
	\label{fig:perfwithsparp100}
\end{figure}

\textbf{Time vs. sparsity}. As we have shown the efficiency of GCOOSpDM in large range of matrices and sparsity, we want to study further about the performance related to the sparsity $s$. We take two matrices with medium ($n=4000$) and large ($n=14000$) dimensions to show the relationship between performance and sparsity. The range of sparsity is kept at $[0.95, 0.9995]$. Here we also put the time cost of the dense algorithm from cuBLAS into comparison so that we can understand under what sparsity GCOOSpDM can outperform cuBLAS. The results for these two sizes of matrices on GTX980, TitanX and P100 GPUs are shown in Fig. \ref{fig:perfwithspar980}, \ref{fig:perfwithspartitanx} and Fig. \ref{fig:perfwithsparp100}, respectively. On one hand, it can be seen that cuBLAS has a constant time cost when the sparsity of matrix increases since the dense algorithm does not consider zero values. On the other hand, the sparse algorithms of cuSPARSE and GCOOSpDM tend to have a linear speedup when the sparsity increases. Given the two specific dimensions of matrices, GCOOSpDM outperforms cuSPARSE with all sparsity. When the sparsity becomes larger than some thresholds, the sparse algorithm would have advantages than the dense one. However, cuSPARSE needs the sparsity be up to $0.995$ to outperform cuBLAS, while our proposed algorithm GCOOSpDM can outperform cuBLAS with sparsity larger than $0.98$. In summary, the GCOOSpDM algorithm is more applicable for matrix multiplication on GPUs than cuSPARSE and cuBLAS under sparsity larger than $0.98$ to achieve higher performance on current GPUs.

\textbf{Performance vs. matrix size}. To further show the sensitivity of the algorithm to the matrix size, we demonstrate the throughput (GFLOPS) in a range of matrix dimensions (i.e., $n\in [400, 14000]$) at two sparsity $0.98$ and $0.995$. The experimental results with sparsity of $0.98$ and $0.995$ are in Fig. \ref{fig:perfwithspar980}, \ref{fig:perfwithspartitanx} and \ref{fig:perfwithsparp100} on three different GPUs. On the three tested GPUs, GCOOSpDM outperforms cuSPARSE with different values of $n$ and two sparsity. For small matrices (e.g., $n<1500$), cuBLAS still outperforms GCOOSpDM since it takes only a small number of cycles in calculating small matrices while GCOOSpDM needs extra overheads on memory allocation and matrix conversion. Given the sparsity of $0.98$ and $n>2000$, GCOOSpDM achieves similar performance as (or slightly better than) cuBLAS. With the sparsity of $0.995$, cuSPARSE achieves close performance with cuBLAS, while GCOOSpDM outperforms cuBLAS up to $2$ times.

\subsection{Breakdown of time costs}
\begin{figure}[!h]
	\centering
	\subfigure[$n=4000$]
	{
		\includegraphics[width=0.48\linewidth]{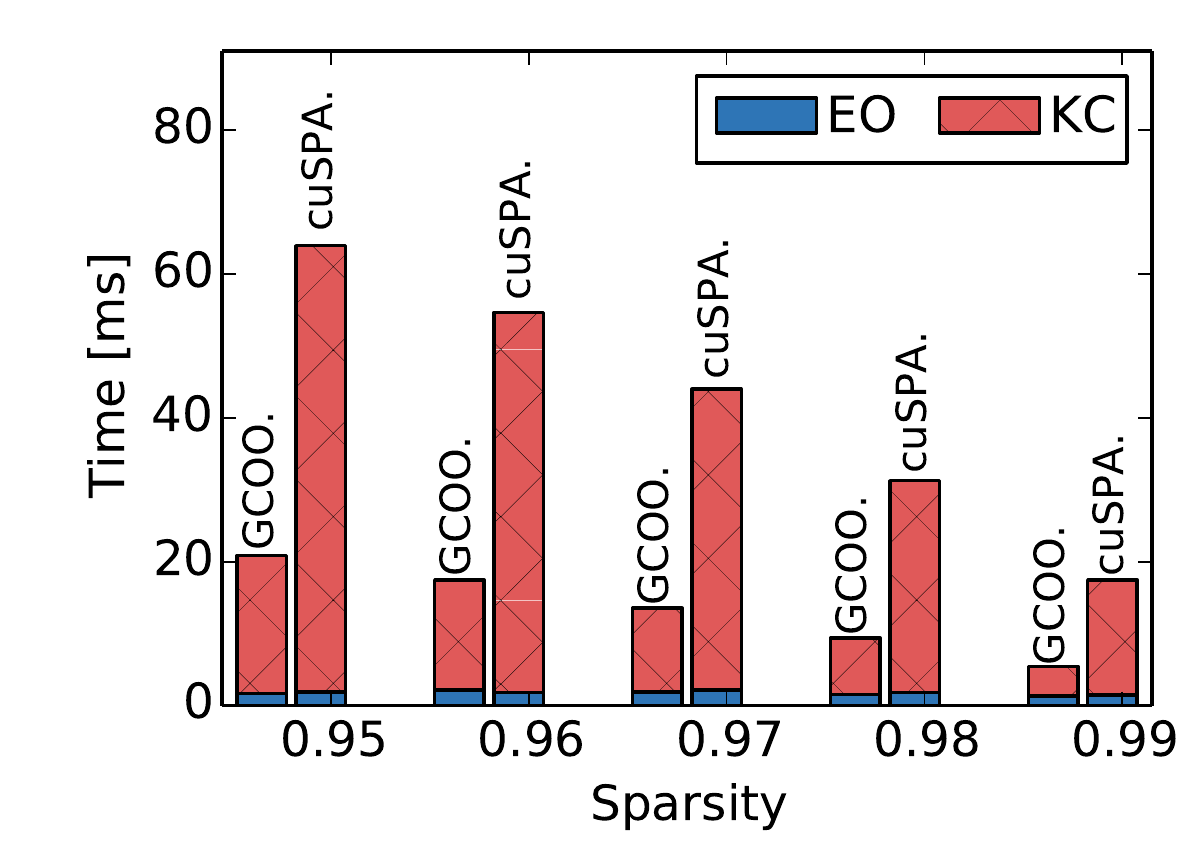}
	}\hspace{-3mm}
	\subfigure[$n=14000$]
	{
		\includegraphics[width=0.48\linewidth]{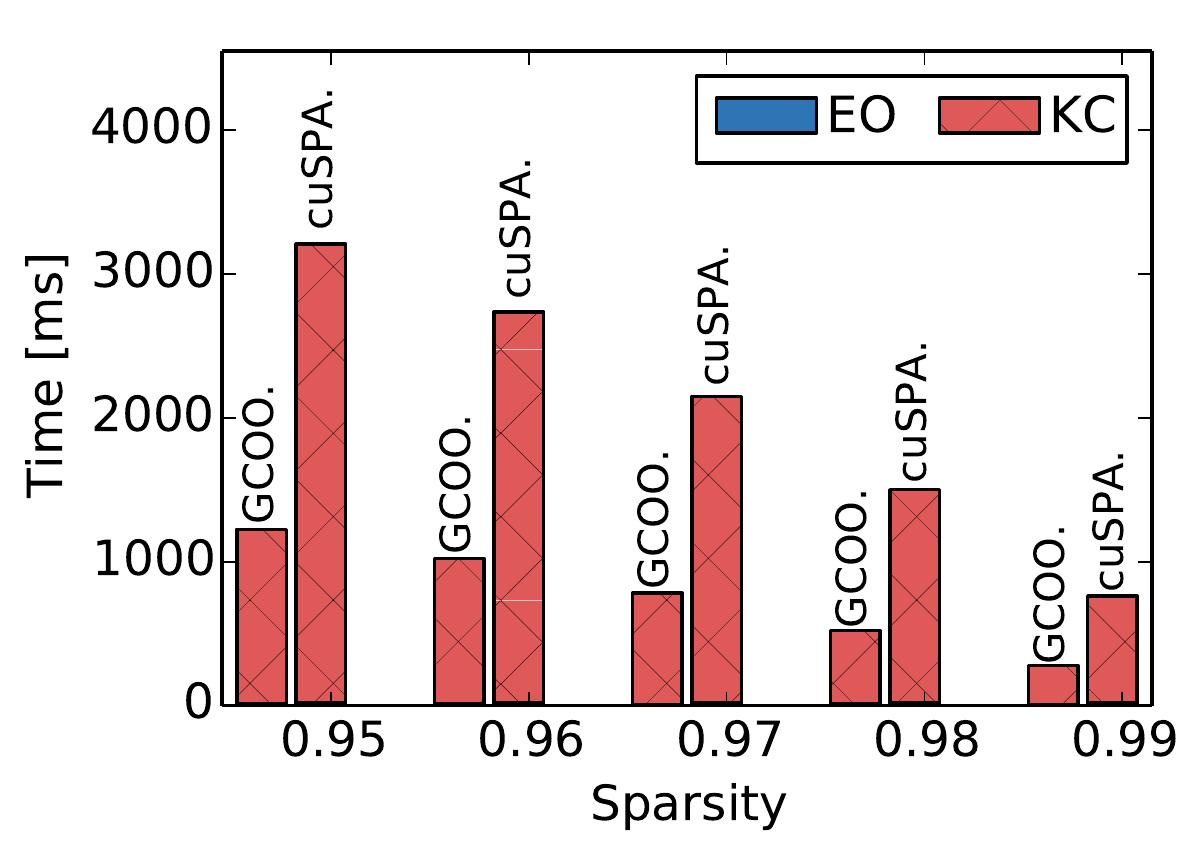}
	}\hspace{-3mm}
	\caption{Time breakdown for two sizes of matrices. ``GCOO.'' represents the GCOOSpDM algorithm, and ``cuSPA.'' represents the algorithm in cuSPARSE.}
	\label{fig:breakdown}
\end{figure}

In this subsection, assume that given $\ma{A}$ and $\ma{B}$ are both in the dense form, while $\ma{A}$ is of high sparsity, we would like to present the time costs of matrix conversion and the kernel calculation to finish the matrix multiplication using the sparse algorithm. The different overheads are summarized into three categories: memory allocation for sparse matrix storage, matrix conversion from the dense form to the sparse form, and SpDM kernel calculation. We summarize the first two categories as an extra overhead (EO), and the third one as the real time cost of kernel calculation (KC). The metrics of EO and KC are used to compare GCOOSpDM and cuSPARSE. Instead of using three GPUs, we only choose a TitanX GPU as our analysis platform, since three GPUs should have similar time distribution. Similar to the previous subsection, we use two sizes of matrices (i.e., $n=4000$ and $n=14000$) with sparsity of $[0.95, 0.96, 0.97, 0.98, 0.99]$ for comparison. The results are shown in Fig. \ref{fig:breakdown}. It can be seen that EO has only a small proportion of the total time, and both GCOOSpDM and cuSPARSE have a very close overhead of EO. The dominated part is the execution time of the kernel that calculates the matrix multiplication.

\subsection{Instruction analysis}
\input{performance_analysis}

\section{Related work}\label{section:relatedwork}
Multiplication of sparse matrices to dense vectors (SpMV) on GPUs have been well studied (e.g., \cite{greathouse2014efficient}\cite{hou2017auto}\cite{bell2009implementing}\cite{merrill2016merge}). Even SpDM can be implemented by multiple SpMVs, the performance could be bad due to a large number of kernel invokes if the matrix is with a large dimension. However, some optimization principles can be applied for SpDM. For example, Yang et al. \cite{yang2018design} use split row \cite{bell2009implementing} and merged path \cite{merrill2016merge} to design SpDM algorithms particularly for tall-skinny matrices. 

Regarding the SpDM algorithm analysis, Greiner et al. \cite{greiner2010complexity} propose an I/O model to interpret the lower bound of efficient serial algorithms. Cache oblivious dense and sparse matrix algorithms are presented by Bader et al. for multi-core CPUs \cite{bader2008cache}. Performance benchmarks \cite{ezouaoui2013performance} are conducted to evaluate the efficiency of different sparse matrix formats for SpDM. Koanantakool et al., \cite{koanantakool2016communication} introduce the communication-avoiding SpDM algorithms that are applied in distributed memory systems. Recent work in designing the row reordering technique to achieve better data temporal locality \cite{jiang2020novel} and the dynamic parameter tuning \cite{parger2020speck} to improve the SpDM performance on GPUs.  

\section{Conclusion and Future Work}\label{section:conclusion}
Sparse-dense matrix-matrix multiplication is commonly used in many scientific computing areas, while designing such algorithms on modern GPUs is non-trivial due to the irregular structure of the sparse matrix. In this paper, we propose an efficient sparse matrix-dense matrix multiplication algorithm on GPUs, called GCOOSpDM. The main optimization techniques used in our algorithm are the coalesced global memory access, proper usage of the shared memory, and reuse the data from the slow global memory. The experimental results show that our proposed algorithm outperforms the vendor-based library: cuSPARSE several times on both the public sparse dataset and randomly generated matrices on three recent Nvidia GPUs (i.e., GTX 980, Titan X Pascal, and Tesla P100). We also analyze in depth the performance improvement on instruction-level to understand why GCOOSpDM performs better than cuSPARSE. The key observation of the instruction-level analysis is that the reduced number of global memory access contributes a lot to the performance gain.

It is difficult for a single algorithm to fit all structures of matrices, sparsity and different types of GPUs. Auto-tune algorithms play an important role for algorithms to find efficient configuration or implementations in different cases. We would like to consider the auto-tune scheme to set proper $p$ and $b$ for our GCOOSpDM algorithm in the future work, and try to extend the GCOO storage format to the multiplication of two sparse matrices.
\bibliographystyle{IEEEtran}
\Urlmuskip=0mu plus 1mu
\bibliography{cites}
\end{document}

%% file: performance_analysis.tex
In this subsection, we compare the instruction distributions of cuSPARSE and GCOOSpDM and explore how the matrix dimension $n$ and the sparsity $s$ take effects on them. The instruction distribution is the runtime statistics of kernel instructions executed on the real GPU hardware. Not only does it help reveal the major performance bottleneck of the GPU kernel, but also determine some quantitative relationships between instructions and kernel performance. 

We use \textit{nvprof}\footnote{\url{http://docs.nvidia.com/cuda/profiler-users-guide}} to collect the runtime instructions of different types, including single-precision floating-point operations, DRAM memory access, L2 cache access, shared memory access and L1/Texture memory access. We use the TitanX GPU as our testbed in the profiling experiments. The other two GPU platforms, GTX980 and P100, can be analyzed with the same experimental methodology. 

\begin{figure}[!h]
	\centering
	\subfigure[cuSPARSE, $s = 0.995$]
	{
		\includegraphics[width=0.45\linewidth]{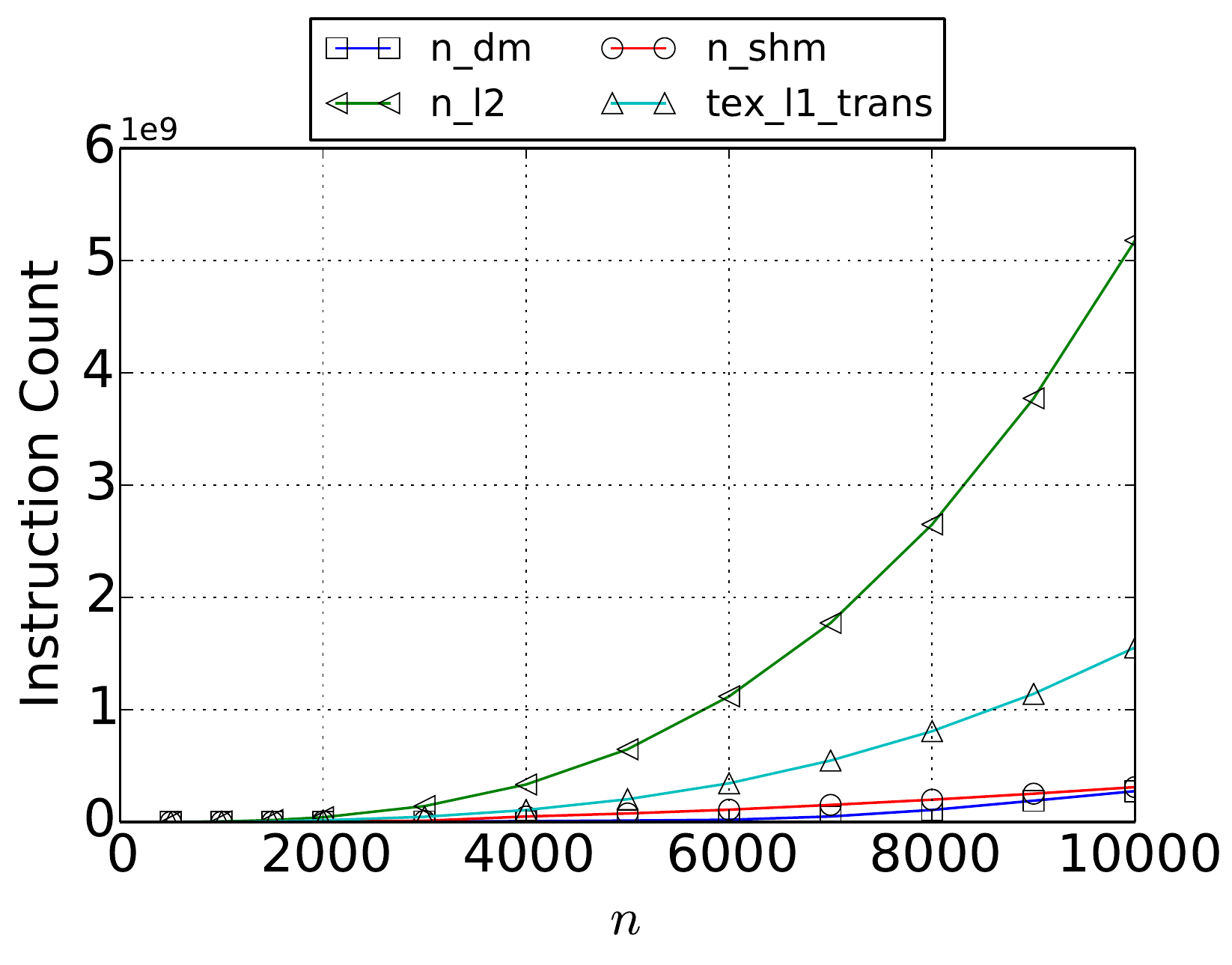}
	}
	\subfigure[GCOOSpDM, $s = 0.995$]
	{
		\includegraphics[width=0.45\linewidth]{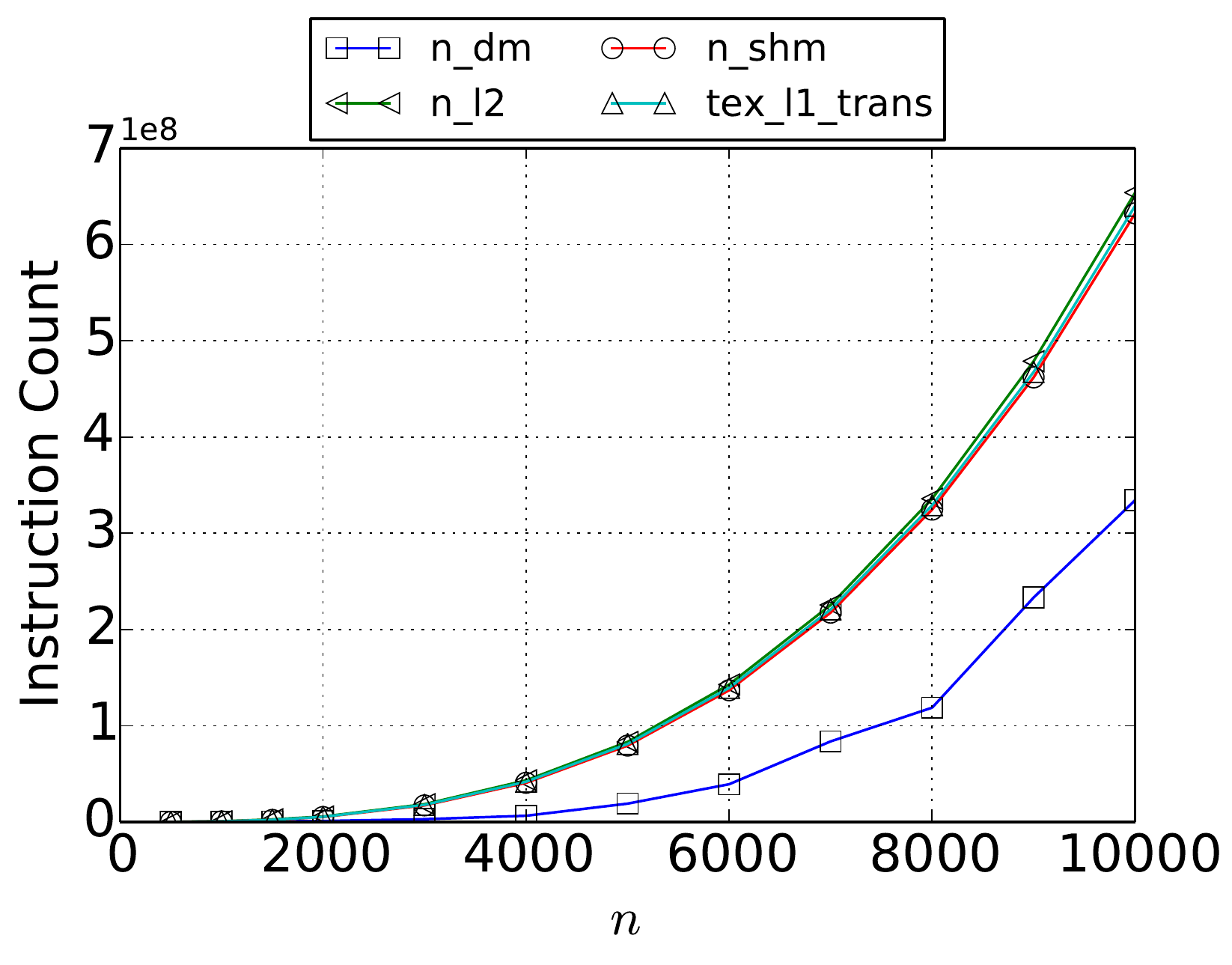}
	}
	\subfigure[cuSPARSE, $n = 4000$]
	{
		\includegraphics[width=0.45\linewidth]{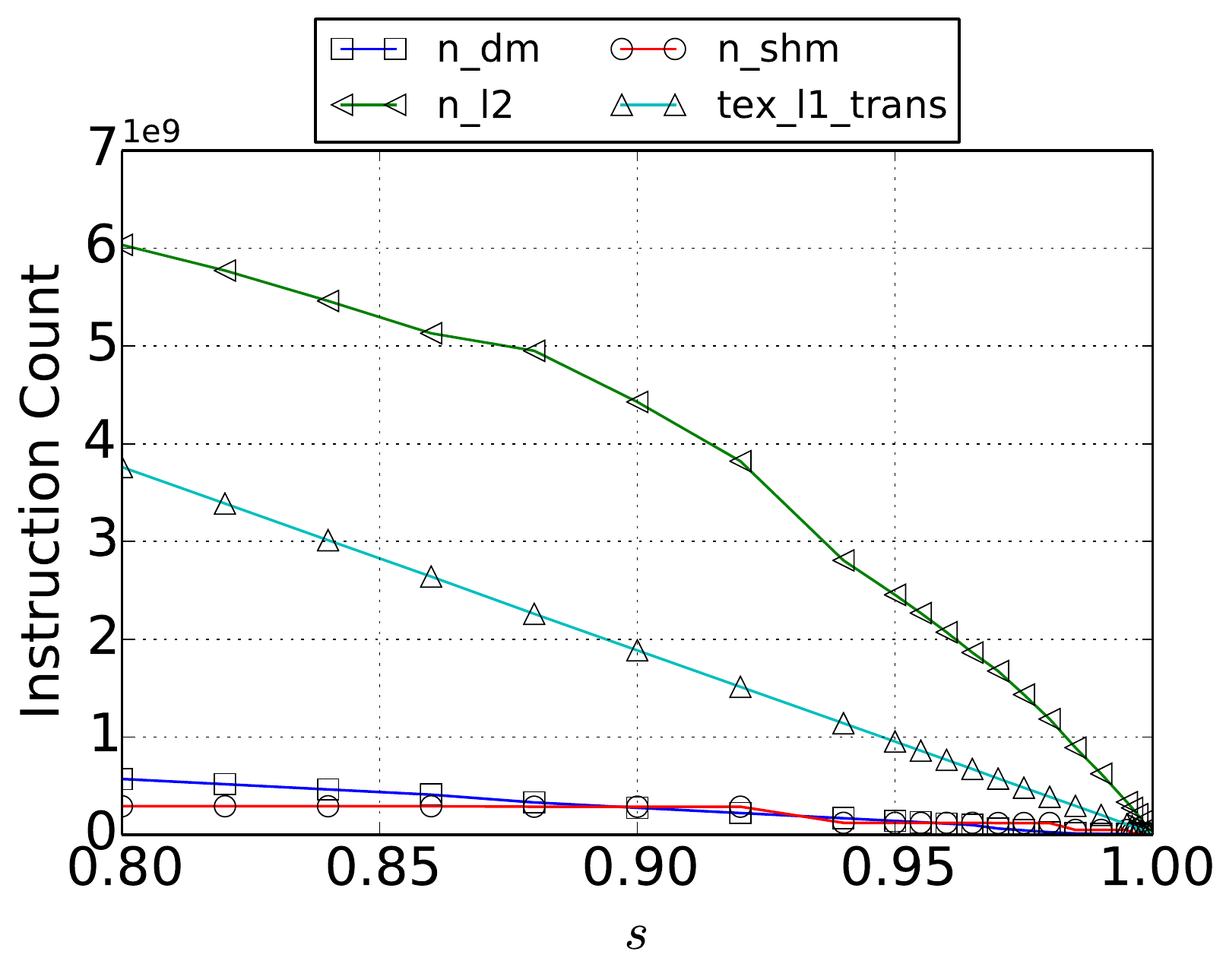}
	}
	\subfigure[GCOOSpDM, $n = 4000$]
	{
		\includegraphics[width=0.45\linewidth]{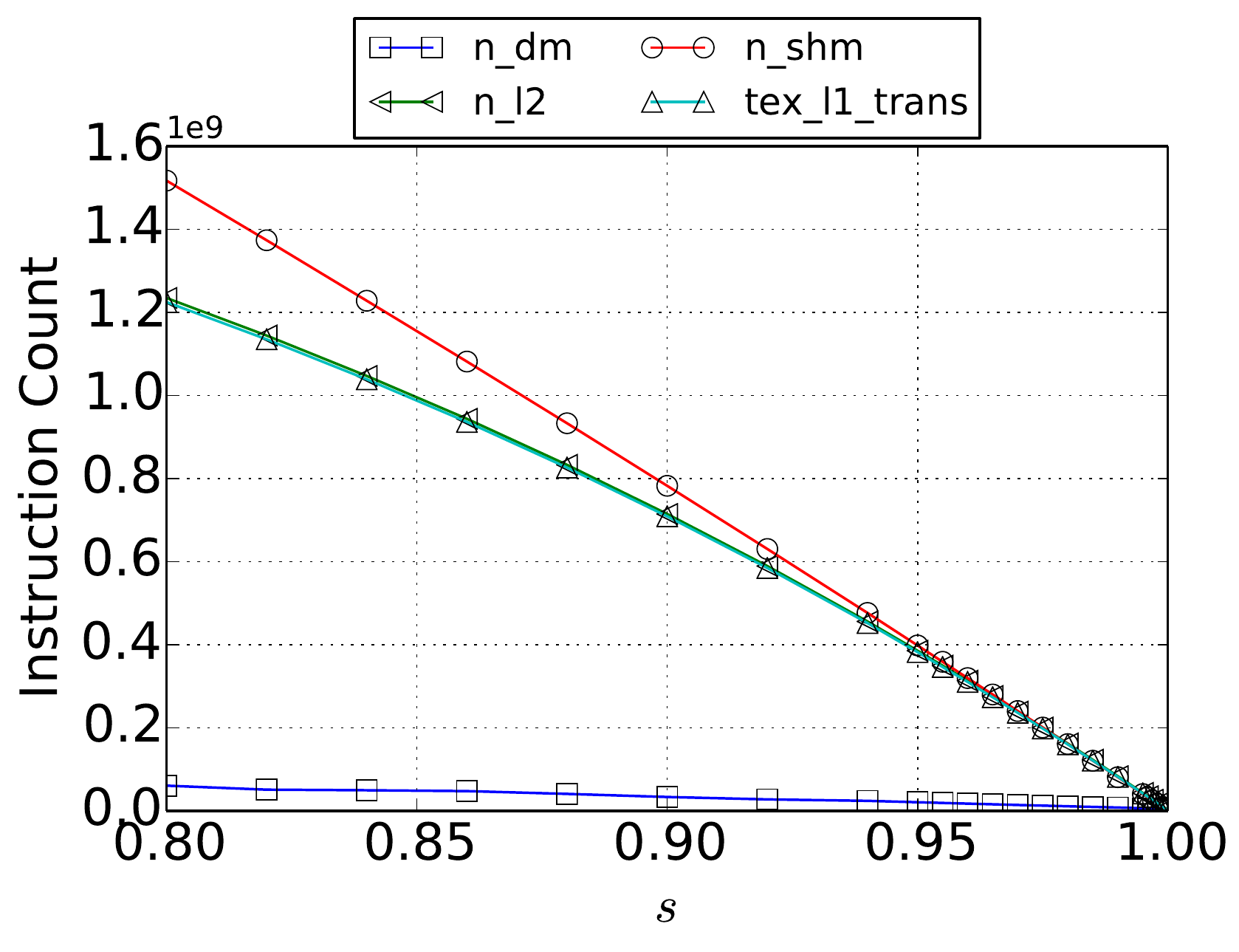}
	}
	\caption{The instruction distribution comparison with respect to the matrix size $n$ and the sparsity $s$ between cuSPARSE and GCOOSpDM on the TitanX GPU. The upper two figures show instruction distributions of different $n$ with fixed $s=0.995$. The bottom two figures show instruction distributions of different $s$ with fixed $n=4000$.}
	\label{fig:analysis_insts}
\end{figure}

We conduct two sets of random sparse matrix experiments on cuSPARSE and GCOOSpDM respectively. First, we fix the matrix sparsity $s$ as $0.995$ and scale the matrix dimension $n$ from $500$ to $10000$. This setting helps exploit how $n$ affects the instructions of those two algorithms. Second, we fix the matrix dimension $n$ as $4000$ and scale the matrix sparsity $s$ from $0.8$ to $0.9995$. This setting helps exploit how $s$ affects the instructions of those two algorithms. Furthermore, we can also witnesses the difference of instruction distributions of cuSPARSE and GCOOSpDM under the same experimental setting. The results are demonstrated in Fig. \ref{fig:analysis_insts}, in which $n\_dm$ denotes the number of DRAM memory access transactions, $n\_l2$ denotes the number of L2 cache access transactions, $n\_shm$ denotes the number of shared memory access transactions, $tex\_l1\_trans$ denotes the number of L1/Texture memory access transactions. We find that the DRAM memory access transactions of both two algorithms only take a very few percentage of total number of memory access transactions. Recall that the DRAM memory has the highest access latency and lowest throughput in the GPU memory hierarchy. Avoidance of very frequent DRAM memory access helps decrease the data fetch overhead of the GPU kernel execution. Both cuSPARSE and GCOOSpDM have well-organized data access patterns to utilize L2 cache and on-chip cache (including shared memory and L1/Texture cache). However, the major parts of memory access instructions of those two algorithms are different. $n\_l2$ takes great majority in cuSPARSE, while $n\_l2$, $n\_shm$ and $tex\_l1\_trans$ take approximately the same partitions in GCOOSpDM. GCOOSpDM has higher utilizations of on-chip cache of GPUs than cuSPARSE so that it generally outperforms cuSPARSE on randomly generated sparse matrices, which confirms the experimental results in Fig. \ref{fig:sythperf}.

\begin{figure}[!h]
	\centering
	\subfigure[Scaling $n$, $s = 0.98$]
	{
		\includegraphics[width=0.45\linewidth]{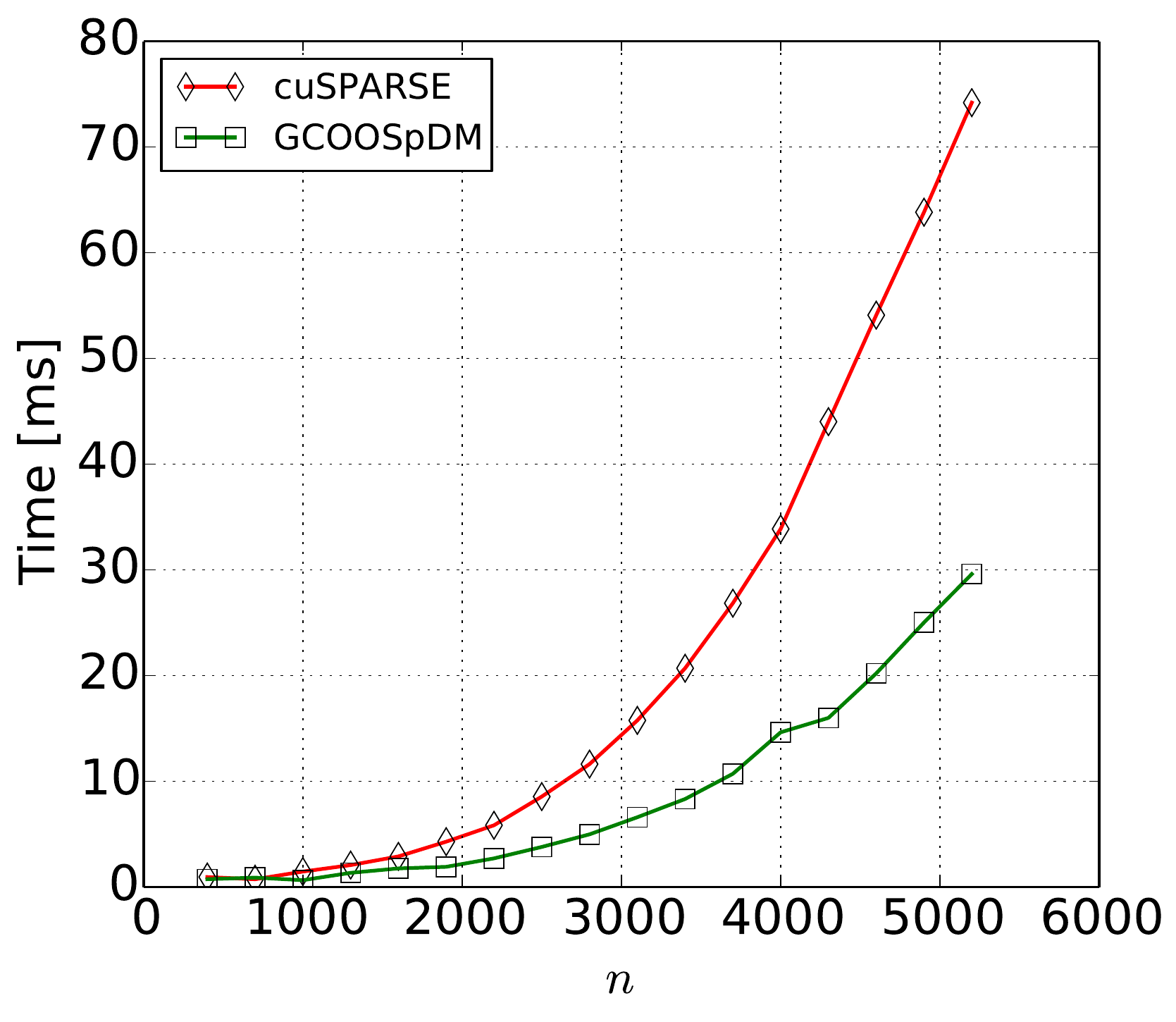}
	}
	\subfigure[Scaling $s$, $n = 4000$]
	{
		\includegraphics[width=0.45\linewidth]{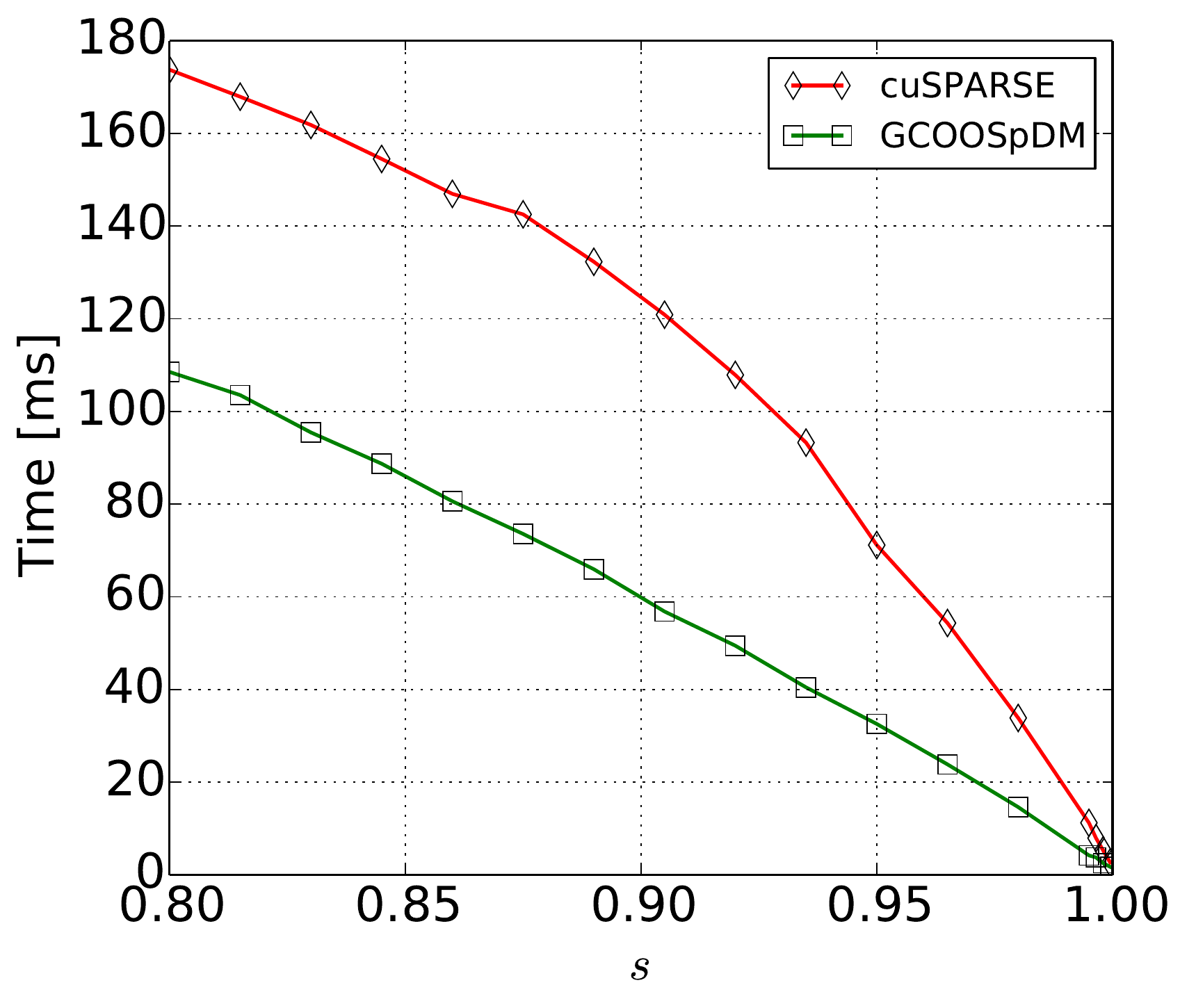}
	}
	\caption{The performance scaling behaviors with respect to the matrix size $n$ and the sparsity $s$ between cuSPARSE and GCOOSpDM on the TitanX GPU. The lower the better.}
	\label{fig:analysis_N_s}
\end{figure}

We then focus on how $n$ and $s$ influence the numbers of those major memory access instructions. The above two figures in Fig. \ref{fig:analysis_insts} show the effects of $n$ on cuSPARSE and GCOOSpDM respectively, while the bottom two show the effects of $s$. We observe that $n\_l2$ of cuSPARSE and $n\_l2$, $n\_shm$ and $tex\_l1\_trans$ of GCOOSpDM all indicate quadratically increasing trends with respect to $n$. It is reasonable since the element number of the output matrix $\ma{C}$ is $n^2$, each of which needs nearly equal workloads of one vector dot product operation. However, the effects of $s$ show a few differences. $n\_l2$ of cuSPARSE performs a nearly quadratically decreasing trend with respect to $s$, while $n\_l2$, $n\_shm$ and $tex\_l1\_trans$ of GCOOSpDM show a nearly linearly decreasing trend. Those observations are also reflected in the performance changing behaviors with respect to $n$ and $s$, as illustrated in Fig. \ref{fig:analysis_N_s}. On one hand, as the matrix size $n$ increases, the performance of both cuSPARSE and GCOOSpDM demonstrates similar quadratically increasing trends, which meets changing behaviors of their dominating memory instructions. On the other hand, as matrix sparsity $s$ increases, the performance of cuSPARSE shows an approximately quadratically decreasing trend, while that of GCOOSpDM shows a linearly decreasing trend. They are also similar to those changing behaviors from exploring the effects of $s$ to the dominating memory instructions of those two algorithms. 

